\documentclass{article}
\usepackage{amsmath}
\usepackage{amssymb}
\usepackage{slashed}
\usepackage[colorlinks=true,bookmarks=false]{hyperref}
\hypersetup{linkcolor=blue,citecolor=blue,urlcolor=blue}
\usepackage{graphicx}
\usepackage[centerlast,small,bf]{caption}

\title{{\huge \textsf{Helicity Spinor Methods and Tree Level QCD Corrections in Higgs Production via Weak Boson Fusion}}}

\author{\textsf{A.R. Fazio and S.C. Vargas}
\\
\textit{Departamento de F\'{\i}sica, Universidad Nacional de Colombia.}
\\
\textit{Ciudad Universitaria, Bogot\'a D.C., Colombia.}
\\
\textit{E-mail:} \texttt{arfazio@unal.edu.co,scvargasa@unal.edu.co}}

\date{}

\begin{document}

\maketitle

\begin{abstract}
{\normalsize
We calculate analytically tree level amplitudes for Higgs production via Weak Bosons Fusion (WBF) and QCD corrections to the Standard Model prediction (SM)  by using the helicity spinor formalism. We provide the explicit expression for the amplitude with  two gluons emission in the final state together with massless final two quarks. The massive generalization of the Britto-Cachazo-Feng-Witten (BCFW) recursion scheme is applied successfully obtaining compact results which agree with, and overcome in simplicity, the conventional Feynman diagrams approach.

}
\end{abstract}

\newpage
\begin{center}
\rule[2mm]{12cm}{1pt}
\end{center}
\tableofcontents
\begin{center}
\rule[2mm]{12cm}{1pt}
\end{center}

\section{Introduction}

Weak Boson Fusion (WBF) process involves promising results in the context of current researches, mainly centered in the Standard Model (SM) examination \cite{zeppenfeld2003,hagiwara}. WBF consists in the generation of a Higgs particle from a weak boson in association with two hard (in the present case, fermion) jets in the forward and backward regions of the detector. It is an important channel in the search for the Higgs boson and a relevant source of data in the measure of its mass, in the study of CP properties and the couplings to the top quark, tau lepton and weak gauge bosons \cite{zeppenfeld2000,zeppenfeld2001,zeppenfeld2004}. More recently, it has been pointed as a relevant channel in the study of the $Hb\bar b$ Yukawa cupling \cite{mangano,asner} and the frontiers of new High Energy Physics \cite{zeppenfeld2006}.

\smallskip
On-shell Recursive formulations based on the helicity spinor formalism have proved to be powerful tools in phenomenology with a far reach. Britto-Cachazo-Feng-Witten (BCFW) \cite{dixon1,dixon2,peskin,dreiner,bern,glover} approach conduces to fast computations of multi-leg partial color-ordered amplitudes that become the essential gauge-invariant constituents of complete amplitudes as a consequence of the basic polar behavior and analyticity properties of the $S$ matrix. Discrete symmetries and the introduction of adequate reference structures based in gauge invariance increase the effectiveness of the scheme making amplitude computation more efficient.

\smallskip
Further schemes had already introduced the characterization of both massive fermions and bosons in this formalism, the expansion to one and two loops calculations in some processes \cite{dixon3} and also its use in numerical schemes. Nevertheless, there are still some features in the path of exploration at tree level with desirable potential in amplitudes estimation. In this context, the study of inverse soft limit process as a solely scheme for amplitudes construction is particularly relevant \cite{boucher} as it is the study of collinear momenta limits.

\smallskip
In previous investigations \cite{figy,ciccolini}, it has been stated the relevance of QCD corrections to WBF process, particularly, additional gluon emissions. In this same spirit, the purpose of this paper is to explore the convenience of introducing the BCFW recursion scheme in this particular context. The objective was to obtain tree level amplitudes with the SM couplings, up to second order in QCD and at first order in WBF Higgs production process. It was considered the limit of massless fermions and the amplitudes were obtained with every external particle as a final state. There were considered up to two gluon emissions and the results agreed with the associated Feynman diagrams in all the cases. In this particular process, while more conventional approaches had been taken into account, the BCFW approach discussed in this paper is original and has not been observed in the literature.

\smallskip
After stating in a explicit way the convention chosen for helicity spinors, polarization vectors and amplitudes (Section \ref{conventions}), the fundamentals of BCFW scheme at tree level are discussed in Section \ref{BCFWFormalism} (specific known results and properties of helicity spinor formalism are presented in Appendix \ref{results}). Once this formalism is introduced, the details of its application in the WBF process are shown in Section \ref{WBFandHiggsProduction}, in which the essential blocks needed for the computation and the way symmetries relate these amplitudes are also discussed. The relevant tree level amplitudes related to WBF process and the studied QCD corrections (1 and 2 gluon emissions) are written explicitly and a final discussion is presented in Section \ref{conclusions}. 

\section{Convention}\label{conventions}

The essential tool of the formalism, the helicity spinor, has been introduced with several conventions. Despite the fact that they are all equivalent, it is easy to get messy. It is then relevant to specify the particular convention and, if possible, state how some of the known identities and results look with it; these are presented in Appendix \ref{results}.

\subsection{Weyl Spinors and Polarization Vectors}

\smallskip
We made use of Weyl spinors, those satisfying massless Dirac equation in $3+1$ dimensions in Minkowski space,
\begin{equation}
\slashed p U\left( p \right) = 0.
\end{equation}
It was considered the signature $g_{\mu\nu}=diag\left(1,-1,-1,-1\right)$ and the following chiral representation for $\gamma$ matrices,
\begin{equation}
{\gamma ^\mu } = \left( {\begin{array}{*{20}{c}}
0&{{\sigma ^\mu }}\\
{{{\bar \sigma }^\mu }}&0
\end{array}} \right),
\end{equation}
\begin{equation}
{\gamma ^5} = \left( {\begin{array}{*{20}{c}}
{ - 1}&0\\
0&1
\end{array}} \right),
\end{equation}
with ${\sigma ^\mu } = \left( {1,\vec \sigma } \right)$ and ${{\bar \sigma }^\mu } = \left( {1, - \vec \sigma } \right)$. We select the right and left 4-component Lorentz irreducible components of massless spinors by
\begin{equation}\label{2compr}
{U_R}\left( p \right) = \frac{1}{2}\left( {1 + {\gamma ^5}} \right)U\left( p \right) = \left( {\begin{array}{*{20}{c}}
0\\
{{u_R}\left( p \right)}
\end{array}} \right),
\end{equation}
\begin{equation}\label{2compl}
{U_L}\left( p \right) = \frac{1}{2}\left( {1 - {\gamma ^5}} \right)U\left( p \right) = \left( {\begin{array}{*{20}{c}}
{{u_L}\left( p \right)}\\
0
\end{array}} \right).
\end{equation}
For massless helicity $\frac{1}{2}$ spinors, the values of helicity and chirality are just proportional. In this case, Lorentz transformations imply for anti-fermions \cite{weinberg}
\begin{equation}
{V_R}\left( p \right) = {U_L}\left( p \right),
\end{equation}
\begin{equation}
{V_L}\left( p \right) = {U_R}\left( p \right).
\end{equation}
By the isomorphism in (\ref{2compr}) and (\ref{2compl}), we can write all the dynamics in Weyl's 2-component formalism as
\begin{equation}
{{\bar U}_L}\left( p \right){U_R}\left( k \right) = U_L^\dag \left( p \right){\gamma ^0}{U_R}\left( k \right) = u_L^\dag \left( p \right){u_R}\left( k \right) \equiv \left\langle {pk} \right\rangle,
\end{equation}
\begin{equation}
{{\bar U}_R}\left( p \right){U_L}\left( k \right) = U_R^\dag \left( p \right){\gamma ^0}{U_L}\left( k \right) = u_R^\dag \left( p \right){u_L}\left( k \right) \equiv \left[ {pk} \right],
\end{equation}
\begin{equation}
{{\bar U}_L}\left( p \right){\gamma ^\mu }{U_L}\left( k \right) = u_L^\dag \left( p \right){{\bar \sigma }^\mu }{u_L}\left( k \right) = \left\langle {\left. {p\left| {{{\bar \sigma }^\mu }} \right|k} \right]} \right.,
\end{equation}
\begin{equation}
{{\bar U}_R}\left( p \right){\gamma ^\mu }{U_R}\left( k \right) = u_R^\dag \left( p \right){\sigma ^\mu }{u_R}\left( k \right) = \left. {\left[ {p\left| {{\sigma ^\mu }} \right|k} \right.} \right\rangle,
\end{equation}
\[ \vdots \]
in which the angle and square spinorial bras and kets are defined.

\smallskip
Observing the behavior under Lorentz group elements, it is possible to choose \cite{peskin2}
\begin{equation}\label{leftright}
{u_R}\left( k \right) = \varepsilon u_L^*\left( k \right),
\end{equation}
with
\begin{equation}
\varepsilon  \equiv i{\sigma ^2} = \left( {\begin{array}{*{20}{c}}
0&1\\
{ - 1}&0
\end{array}} \right).
\end{equation}
More details are presented in Appendix \ref{resultsweyl}.

A polarization vector for a massless final gauge boson with specific helicity and momentum $k$ can be written as
\begin{equation}\label{polr}
\varepsilon _ + ^{*\mu }\left( {k,r} \right) = \frac{1}{{\sqrt 2 }}\frac{{\left\langle {\left. {r\left| {{{\bar \sigma }^\mu }} \right|k} \right]} \right.}}{{\left\langle {rk} \right\rangle }},
\end{equation}
\begin{equation}\label{poll}
\varepsilon _ - ^{*\mu }\left( {k,r} \right) =  - \frac{1}{{\sqrt 2 }}\frac{{\left. {\left[ {r\left| {{\sigma ^\mu }} \right|k} \right.} \right\rangle }}{{\left[ {rk} \right]}},
\end{equation}
with $r$ a reference massless vector not collinear with $k$. The definitions (\ref{polr}) and (\ref{poll}) satisfy all the requirements for polarization vectors (see Appendix \ref{resultspolvec}), as wave functions of massless gauge bosons of a Lorentz invariant Quantum Field Theory.

\subsection{Amplitudes}\label{convampl}

All the momenta will be considered outgoing; all the particles will be treated as final states. Initial state particles may be recovered by crossing. The outgoing right- and left-handed fermions $\left( {j_f^ + ,j_f^ - } \right)$ will be represented with the 2-component spinors $u_R^\dag \left( {{p_j}} \right),u_L^\dag \left( {{p_j}} \right)$ (associated with the 4-component spinors ${\bar U_R}\left( {{p_j}} \right),{\bar U_L}\left( {{p_j}} \right)$). Outgoing right- and left- handed antifermions $\left( {j_{\bar f}^ + ,j_{\bar f}^ - } \right)$ will be represented with the 2-component spinors ${u_L}\left( {{p_j}} \right),{u_R}\left( {{p_j}} \right)$ (which are associated with the 4-component spinors ${U_L}\left( {{p_j}} \right) = {V_R}\left( {{p_j}} \right),{U_R}\left( {{p_j}} \right) = {V_L}\left( {{p_j}} \right)$).

\smallskip
There are some additional observations. The explicit *'s on the polarization vectors will be dropped. We will rescale SU(3) generators as ${T^a} \equiv \sqrt 2 {t^a}$, so the normalization convention becomes $Tr\left[ {{T^a}{T^b}} \right] = {\delta ^{ab}}$ and also QCD Feynman Rules are written in terms of the new ${T^a}$ operators. And finally, in massive gauge boson propagators it is introduced the complex term related to the width ${\Gamma _V}$ of the respective particle.

\smallskip
Since we'll be considering the Standard Model couplings, it is natural to introduce the conventional helicity coupling factors. In the present case, they are going to be used only the following couplings between fermions and gauge vector bosons (indicated with $V$), ${h_{cf,V,AB}}\left( {hel} \right)$, given by
\begin{equation}\label{couplingWplus}
{h_{cf,W,AB}}\left(  +  \right) = 0,
\end{equation}
\begin{equation}\label{couplingWminus}
{h_{cf,W,AB}}\left(  -  \right) = \frac{1}{{\sqrt 2 sin\left( {{\theta _w}} \right)}}{U_{CD}},
\end{equation}
\begin{equation}\label{couplingZplus}
{h_{cf,Z,q}}\left(  +  \right) = \frac{{ - 2{Q_q}si{n^2}\left( {{\theta _w}} \right)}}{{sin\left( {2{\theta _w}} \right)}},
\end{equation}
\begin{equation}\label{couplingZminus}
{h_{cf,Z,q}}\left(  -  \right) = \frac{{ \pm 1 - 2{Q_q}si{n^2}\left( {{\theta _w}} \right)}}{{sin\left( {2{\theta _w}} \right)}},
\end{equation}
with $hel$ the helicity in the ${U^{hel}}\left( p \right)$ spinor linked with the vertex, $C$ ($D$) being the up (down) type quark between $A$ and $B$, ${U_{CD}}$ the CKM Matrix, and the two signs, $\pm$, corresponding to up ($+$) and down ($-$) type quarks. The remaining notation is conventional: ${{\theta _w}}$ is the Weinberg angle and ${{Q_q}}$ the charge of the quark. Also, it is convenient to make use of the couplings of the vector bosons with the Higgs scalar, ${H_{cf,V}}$,
\begin{equation}\label{couplingWHiggs}
{H_{cf,W}} = cot\left( {{\theta _w}} \right),
\end{equation}
\begin{equation}\label{couplingZHiggs}
{H_{cf,Z}} = \frac{2}{{sin\left( {2{\theta _w}} \right)}}.
\end{equation}
The corresponding Feynman rules used in the following computations are just the product of the previous couplings, an $i$ factor and the associated coupling constants ($e$ in the cases (\ref{couplingWplus}) to (\ref{couplingZminus}), and $eM_z$ in the cases (\ref{couplingWHiggs}) and (\ref{couplingZHiggs})), \cite{bern}. QCD rules will be discussed in Sections \ref{BCFWobsandconv} and \ref{basic}.

\section{BCFW On Shell Recursion Relations at Tree Level}\label{BCFWFormalism}

Consider a color-ordered partial amplitude $iM\left( {k_1^{{h_1}},...,k_m^{{h_m}},p_1^{{h_1}},...,p_n^{{h_n}}} \right)$, in which the colored particles $\left( {p_1^{{h_1}},...,p_n^{{h_n}}} \right)$ come in a definite cyclic order $1,...,n$, and some colorless particles $\left( {k_1^{{h_1}},...,k_m^{{h_m}}} \right)$ are produced too (for instance, one Higgs scalar in the present case of interest). With partial it is meant the complete amplitude with the color factors stripped away. Therefore, it depends on the momenta and helicities only. It is going to be treated the tree level case.

\smallskip
Now, choose two particles in the amplitude. Their spinors are then shifted by a complex vector. This shift depends on whether these particles are massless or not. In this case, the chosen particles will be massless fermions, labeled $i$ and $j$. The shift to be used is
\begin{equation}
\left| {\hat i} \right\rangle  = \left| i \right\rangle,
\end{equation}
\begin{equation}
\left| {\left. {\hat i} \right]} \right. = \left| {\left. i \right]} \right. + z\left| {\left. j \right]} \right.,
\end{equation}
\begin{equation}
\left| {\left. {\hat j} \right]} \right. = \left| {\left. j \right]} \right.,
\end{equation}
\begin{equation}
\left| {\hat j} \right\rangle  = \left| j \right\rangle  - z\left| i \right\rangle.
\end{equation}
With the use of Gordon and Fierz identities, one easily finds that
\begin{equation}
\hat p_i^\mu \left( z \right) = p_i^\mu  + \frac{z}{2}\left\langle {\left. {i\left| {{{\bar \sigma }^\mu }} \right|j} \right]} \right.,
\end{equation}
\begin{equation}
\hat p_j^\mu \left( z \right) = p_j^\mu  - \frac{z}{2}\left\langle {\left. {i\left| {{{\bar \sigma }^\mu }} \right|j} \right]} \right.,
\end{equation}
so the momentum is still conserved and the shifted particles remain on-shell, $\hat p_i + \hat p_j = p_i + p_j$ and $\hat p_j^2=0=\hat p_i^2$.

\smallskip
Hence, it is an on-shell scattering amplitude $iM\left( z \right)$ of particles with complex momenta and can be computed from the usual Feynman Rules. Momentum conservation suggests that both the momenta of external particles and the spinors of massless external particles are linear functions of $z$. As a consequence, at tree level $iM\left( z \right)$ is a rational function of $z$. This can be seen as a particular way of continuing a scattering amplitude in the complex plane of momenta, by exploiting the analyticity properties of the $S$-matrix.

\begin{figure}
  \centering
    \includegraphics[width=0.7\textwidth,viewport=5 60 350 300,clip]{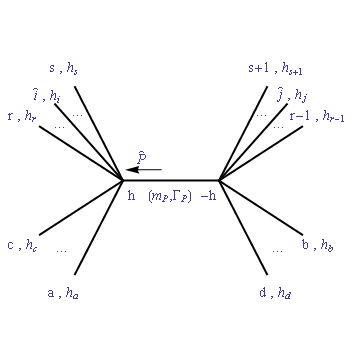}
  \caption{BCFW diagram that illustrates the kind of factorizations found when working on an amplitude. Following the labels introduced for the present case, in this diagram $\left\{ {a,...,c} \right\} \subset \left\{ {1,...,m} \right\}$, $\left\{ {b,...,d} \right\}$ is the set difference of the complete set $\left\{ {1,...,m} \right\}$ and the subset $\left\{ {a,...,c} \right\}$, and  $\left\{ {r,...,i,...,s,s + 1,...,j,...,r - 1} \right\}$ is a cyclic permutation of $\left\{ {1,...,n} \right\}$.}
  \label{BCFW}
\end{figure}

\smallskip
Considering the structure of Feynman Rules and the construction of the shift, at tree level $iM\left( z \right)$ can only have simple poles in $z$ coming from internal propagators, as could be expected from the perspective of the Polology Theorem \cite{weinberg}. In fact, for generic external momenta, all internal momenta are different producing a distinct location for each pole. Also, the residues of these poles are given by all the intermediate states in which the interchanged momentum depends on $z$. Each one of these states represents a factorization between two groups of sub-amplitudes; lets say left and right. According to the shift, the propagator will depend on $z$ only if $i$ and $j$ particles belong to different groups. We choose the diagrams in which the particle $i$ is on the left and the particle $j$ is on the right (clearly, the remaining diagrams repeat the same pole contributions). In Figure~\ref{BCFW}, it is shown a scheme exemplifying these factorizations. Both the location of $i$ and $j$ particles and the constrain on the cyclic order of colored particles is stated explicitly.

\smallskip 
 According to Figure~\ref{BCFW} (see the convention chosen for the direction of the momentum $\hat P$ in the propagator) and the shifts, with the definitions
\begin{equation}
{p_{Left}} \equiv \sum\limits_{i = r}^s {{p_i}},
\end{equation}
\begin{equation}
{k_{Left}} \equiv \sum\limits_{l \in \left\{ {a,...,c} \right\}} {{k_l}},
\end{equation}
\begin{equation}
{P_{Left}} \equiv {p_{Left}} + {k_{Left}},
\end{equation}
one may write
\begin{equation}
{{\hat P}^\mu }\left( z \right) = P_{Left}^\mu  + \frac{z}{2}\left\langle {\left. {i\left| {{{\bar \sigma }^\mu }} \right|j} \right]} \right.,
\end{equation}
\begin{equation}
{{\hat P}^2}\left( z \right) = P_{Left}^2 + z\left\langle {\left. {i\left| {{{\bar P}_{Left}}} \right|j} \right]} \right.
\end{equation}
and
\begin{equation}
\tilde P\left( z \right) \equiv {{\hat P}^\mu }\left( z \right){{\bar \sigma }_\mu } = {{\bar P}_{Left}} + z\left| i \right\rangle \left. {\left[ j \right.} \right|.
\end{equation}

\smallskip
So, any given factorization of this kind, writing explicitly the denominator of the propagator, may be written as (the width ${{\Gamma _P}}$ is introduced if it is the case)
\begin{equation}
\frac{{c\left( z \right)}}{{{{\hat P}^2} - m_P^2 + i{\Gamma _P}{m_P}}} = \frac{{c\left( z \right)}}{{\left\langle {\left. {i\left| {{{\bar P}_{Left}}} \right|j} \right]} \right.\left\{ {z - {z_{part}}} \right\}}},
\end{equation}
with ${m_P}$ the mass of the intermediate particle and
\begin{equation}
{z_{part}} =  - \frac{{\left[ {P_{Left}^2 - m_P^2 + i{\Gamma _P}{m_P}} \right]}}{{\left\langle i \right|{{\bar P}_{Left}}\left| {\left. j \right]} \right.}}.
\end{equation}

\smallskip
Now, if $\mathop {\lim }\limits_{\left| z \right| \to \infty } iM\left( z \right) = 0$, then by integrating over a closed curve ${\rm{\gamma }}$   sufficiently far from the origin, it is found 
\begin{equation}
\frac{1}{{2\pi i}}\oint_{\rm{\gamma }} {\left[ {\frac{{iM\left( z \right)}}{z}} \right] = } \mathop {{\rm{Res}}}\limits_{z = 0} \left[ {\frac{{iM\left( z \right)}}{z}} \right] + \sum\limits_{{z_{poles}} \ne 0} {\mathop {{\rm{Res}}}\limits_{z = {z_{poles}}} \left[ {\frac{{iM\left( z \right)}}{z}} \right]}  = 0,
\end{equation}
then
\begin{equation}
iM\left( 0 \right) =  - \sum\limits_{{z_{poles}} \ne 0} {\frac{{c\left( z_{part} \right)}}{{{z_{part}}\left\langle {\left. {i\left| {{{\bar P}_{Left}}} \right|j} \right]} \right.}}}.
\end{equation}
This is just the sum over all possible factorizations in which the intermediate state momentum depends on $z$ and is equivalent to the sum over all the BCFW diagrams with the structure of Figure~\ref{BCFW}. More explicitly,
\begin{multline}\label{BCFWform}
iM\left( {k_1^{{h_1}},...,k_m^{{h_m}},p_1^{{h_1}},...,p_n^{{h_n}}} \right) = \\ \sum\limits_\omega  {} \sum\limits_\varpi  {} iM\left( {k_a^{{h_a}},...,k_c^{{h_c}},p_r^{{h_r}},...,\hat p_i^{{h_i}},...,p_s^{{h_s}}, - {{\hat P}^h}} \right) \times \\ \times \frac{i}{{P_{Left}^2 - m_P^2 + i{\Gamma _P}{m_P}}}{\left. {iM\left( {k_b^{{h_b}},...,k_d^{{h_d}},{{\hat P}^{ - h}},p_{s + 1}^{{h_s}},...,\hat p_j^{{h_1}},...,p_{r - 1}^{{h_{r - 1}}}} \right)} \right|_{z = {z_{part}}}},
\end{multline}
where the sum over $\omega$ should be understood as a sum over the partitions with $i$ on the left and $j$ on the right that keep the cyclic order in the set of colored particles, and the sum over $\varpi$ as a sum over the intermediate helicities or spin states.

\subsection{Observations and Conventions}\label{BCFWobsandconv}
There are some important observations to point out concerning this result.

\smallskip
Since we'll be working with partial amplitudes, all the color factors and the couplings (helicity and not helicity dependent) are stripped away. In our convention, only the $i$ or $- i$ factors from the vertices and propagators are carried in the partial amplitudes. Also, since the amplitudes are color-ordered, the QCD vertices come from the color-ordered Feynman rules, which are readily obtained from the usual couplings. The process is stated clearly in \cite{peskin} in which the rules in \cite{peskin2} are taken as the starting point. It will be discussed briefly in Section \ref{basic}. These, also, must be constructed according to the prescription of outgoing momenta (in our convention) and making use of the specific cyclic order chosen for the amplitudes. In the present case, the particle label was chosen to grow in the clockwise direction, as can be seen in the used BCFW diagrams. These two prescriptions agree with the ones in \cite{peskin}.

\smallskip
For the present choice made in the shift, it has been found that when the helicities are $\left( {{h_i},{h_j}} \right) \in \left\{ {\left( { + , + } \right),\left( { - , - } \right),\left( { - , + } \right)} \right\}$ , the amplitude shows the desired behavior, $\mathop {\lim }\limits_{\left| z \right| \to \infty } iM\left( z \right) = 0$ \cite{peskin}. Also, there are some restrictions in the kind of particles that can be labeled as $i$ and $j$. For instance, if the two marked particles are a quark and anti-quark of the same flavor, they should not be adjacent. Also, if one chooses a quark and a gluon which are adjacent and share the same helicity, then the order used (with our convention) was $(i,j)=(q^{+},g^{+})$ or $(i,j)=(g^{-},q^{-})$ \cite{glover}.

\smallskip
As can be seen in (\ref{BCFWform}), one must set a convention for the spinors $\left| { - \hat P} \right\rangle$ and $\left| {\left. { - \hat P} \right]} \right.$. It can be written $\left| { - \hat P} \right\rangle  = {e^{i\alpha }}\left| {\hat P} \right\rangle$ and $\left| {\left. { - \hat P} \right]} \right. = {e^{i\beta }}\left| {\left. {\hat P} \right]} \right.$, with $\alpha$ and $\beta$ real phases. By obtaining a known amplitude with no fermion propagators, it is possible to show that one must demand $2\alpha ,2\beta  \in \left\{ {\pi , - \pi } \right\}$. Expressions like $\left| { - \hat P} \right\rangle \left. {\left[ { - \hat P} \right.} \right|$ may be treated as $ - {{\hat P}_\mu }{{\bar \sigma }^\mu } =  - \left| {\hat P} \right\rangle \left. {\left[ {\hat P} \right.} \right|$; this may be considered an extension of the process of analytic continuation made from the positive energy case to introduce negative energy spinors \cite{dixon1}. Additionally, according to the possible phases, in the boson propagator case one may eliminate the minus signs in the momenta of the spinors and insert a minus sign for each pair of these spinors of the same kind (angle or square) in a product (if it is a quotient, it is clear that the phases vanish and no minus is introduced).

\smallskip
From the operational definition proposed for left and right spinors, it may be expected one phase, ($\alpha $ or $\beta $) to be the opposite of the other one. However, as pointed out in \cite{dixon2}, for complex momenta, angle and square spinors are not in general complex conjugates of each other. Observe that if ${p_j} \ne p_j^*$, then
\[{\left( {\left| j \right\rangle \left. {\left[ j \right.} \right|} \right)^\dag } = {\left( {{p_j}_\mu {{\bar \sigma }^\mu }} \right)^\dag } = p_{j\mu }^*{{\bar \sigma }^\mu } \ne {p_j}_\mu {{\bar \sigma }^\mu } = \left| j \right\rangle \left. {\left[ j \right.} \right|.\]
In particular, this permits to introduce nonsingular tree level amplitudes for 3 legs vertices in which all momentum invariants vanish \cite{dixon2}.

\smallskip
It is also necessary to make a correction when dealing with fermion propagators. Observe that in the BCFW recursion formula the direction in the momentum on the propagator was fixed. As is seen in Figure~\ref{BCFW}, $\hat P$ is shown as an incoming momentum in the left sub-amplitude (and as an outgoing momentum in the right one) and the signs in (\ref{BCFWform}) are stated according to this choice (in order to preserve the prescription of outgoing momenta in partial amplitudes). Because of this, additional multiplicative phases must be introduced according to the real sign in the slash factor that would be expected for the propagator in a Feynman amplitude.

\smallskip
In order to understand how these phases are introduced, there are some important observations to be made. First, it can be shown that, in general, an intermediate fermion state appears with a factorization of amplitudes in which the left one has the angle or square ket with momentum $- \hat P$ an odd number of times. This can be seen as a consequence of the BCFW process since elaborate amplitudes are worked out from simpler ones and they are all constructed from a set of basic blocks which come directly from the Feynman vertices (see Section \ref{basic}). In those blocks, the kets or bras associated with boson particles appear an even number of times while the fermion particles appear an odd number of times. As the BCFW scheme is applied, these characteristics are preserved in more complex amplitudes in a inductive process. The denominator of a propagator will always provide an even number of spinors of each particle involved since it is related with the square of the intermediate momentum. The spinors with the intermediate shifted momentum ($- \hat P$ or $\hat P$) will always appear an even number of times when counting both the left and right amplitudes; these are always replaced and written in terms of the known momenta. The expression for ${z_{part}}$ guarantees that $\tilde P\left( {z_{part}} \right) = {{\bar P}_{Left}} + {z_{part}}\left| i \right\rangle \left. {\left[ j \right.} \right|$ always introduces an even number of spinors of each involved momentum. Then, the number of spinors remains odd for fermions and even for bosons after the process is done to construct more complex amplitudes.

\smallskip
Then, in the case of a fermion propagator, after transforming the $- \hat P$ kets in the left amplitude by pairs with the prescription stated before, there will be a remaining ket (since there are an odd number of them) for which the phase is introduced according to the momentum flow in the propagator. Consider the case in which the remaining ket is in the numerator of the amplitude. Then, if the fermion arrow goes in the direction of  $\hat P$, a common Feynman amplitude would demand the introduction of a ${{\hat P}^\mu }\left( z_{part} \right){{\gamma }_\mu } = \left. {\left| \hat P \right.} \right]\left\langle {\left. \hat P \right|} \right. \otimes \left| \hat P \right\rangle \left. {\left[ \hat P \right.} \right|$, so the ket is replaced with no phase, $\left| {\left. { - \hat P} \right]} \right.,\left| { - \hat P} \right\rangle  \to \left| {\left. {\hat P} \right]} \right.,\left| {\hat P} \right\rangle $. If the fermion arrow goes in the opposite direction, it would be associated with a ${{- \hat P}^\mu }\left( z_{part} \right){{\gamma }_\mu }$, so one makes the replacement with a minus, $\left| {\left. { - \hat P} \right]} \right.,\left| { - \hat P} \right\rangle  \to  - \left| {\left. {\hat P} \right]} \right., - \left| {\hat P} \right\rangle $. The case in which the remaining ket is in the denominator results in a minus when the arrow and the momentum directions match, and no phase in the opposite case.

\smallskip
In a computational perspective, this set of prescriptions may be implemented in a more simple and equivalent way by introducing a $i$ factor for each angle or square ket with momentum $- \hat P$ and, in the case of fermion propagators, introducing an additional $- i$ factor if the arrow and the momentum directions match, and a $i$ in the opposite case.

\smallskip
The sum over the intermediate particle states is more often done over the helicities, following the original scheme. In \cite{glover} it was proposed as an alternative the sum over spin states, particularly when dealing with massive intermediate particles. In our case of interest, for instance, this scheme was used when dealing with vector boson propagators and is of special effectiveness when it is possible to use any form of the Ward-Takahashi identity.

\smallskip
As pointed out in \cite{dixon1}, the number of required amplitudes is reduced by a couple of factors. The first one is a number of symmetries that relate amplitudes with different sets of helicities, improving the effectiveness of the helicity approach. Some of these will be discussed in the next sections. In second place, the freedom in the polarization vectors and the forms that vector contractions take (see Appendix \ref{resultspolvec}) guarantees that a lot of QCD amplitudes vanish when the number of gluons with the same helicity is too high. The argument is simple and is treated in several references like \cite{dixon1,peskin}. This is the origin of the designation of the maximally helicity violating or MHV amplitudes; those that hold the maximum number of helicities of the same kind without vanishing. These two features also lead to an even faster application of BCFW formula as the needed blocks became manifest by considering these aspects.

\smallskip
As a final observation, it is well known that in the amplitudes simplifications it is useful to form complete operators with the structures $\left| {\hat P} \right\rangle \left. {\left[ {\hat P} \right.} \right|$ and $\left| {\left. {\hat P} \right]} \right.\left\langle {\hat P} \right|$. This tends to simplify the algebra by contracting with the appropriate factors and reminding that $\left| {\hat P} \right\rangle \left. {\left[ {\hat P} \right.} \right| = {\bar P_{Left}} + {z_{part}}\left| i \right\rangle \left. {\left[ j \right.} \right|$.

\section{WBF and Higgs Production}\label{WBFandHiggsProduction}

The relevant processes in which Higgs particles are expected at the proton-proton LHC collider are gluon fusions ($gg \to H$), WBF ($qq \to qq + {W^*}{W^*},{Z^*}{Z^*} \to qq + H$), Higgs-strahlung ($gg \to W,Z \to W,Z + H$) and Higgs bremsstrahlung off a top quark ($qq,gg \to t\bar t + H$) \cite{straessner}. Despite the fact that gluon fusion process has a higher total cross-section than WBF, it plays an important role in the determination of Higgs couplings and even bounds on non-standard couplings \cite{ciccolini}.

\smallskip
At tree level, WBF process is a pure Electroweak (EW) process, in which the main contributions come from \textit{t} and \textit{u} types of channels as some strong interactions process are suppressed \cite{ciccolini}. NLO QCD corrections are relevant in order to obtain more reliable predictions in order to reduce substantial theoretical uncertainties \cite{figy}. It has been found \cite{figy,ciccolini} that these corrections to integrated cross sections are small but a suppression of virtual gluon corrections observed at one loop is an interesting feature to be examined at higher orders. Recursion schemes may provide a useful approach in this research.

\smallskip
The following computations are obtained with the aid of the helicity spinor formalism and the BCFW tree level recursion scheme with a parallel presentation of the Feynman associated amplitudes. In order to do that, we first introduce some basic blocks. With the SM couplings and up to second order in QCD and at first order in WBF Higgs production process, we obtain tree level amplitudes up to two gluon emissions. Thus, they will be obtained up to second order in weak boson couplings with fermions and up to first order in weak boson couplings with the Higgs scalar. As it was stated before, this BCFW approach has not been observed in the literature for this particular process.

\subsection{BCFW Basic Blocks}\label{basic}

The main blocks needed in the process and its QCD corrections are those corresponding to the vertex between a vector boson and fermions, and the three legs QCD vertices. These blocks are obtained by contracting the vertices with the associated on-shell external legs. As a benefit of the use of the BCFW and due to the non Abelian Ward identity, the four gluons vertex of QCD is not necessary as a basic block.

\smallskip
These basic blocks are obtained directly from their associated color-ordered Feynman rules with the prescriptions chosen for the partial amplitudes pointed out in the previous section. As the matter of fact, it was found that with the right $i$ and $j$ choices, the three gluons vertex was not necessary in the amplitudes we wanted to find. It's remarkable that, in more standard approaches, the three gluons interaction is used as a basic constituent.

\subsubsection{$q{V_\mu }\bar q$ or $Q{V_\mu }\bar q$ Blocks}
According to the conventions stated for partial amplitudes, the non-vanishing blocks are
\begin{equation}\label{qVqbar1}
i{M_\mu }\left( {1_q^ - ,2_{\bar q}^ + } \right) = i\left. {\left[ {2\left| {{\sigma _\mu }} \right|1} \right.} \right\rangle  = i\left\langle {\left. {1\left| {{{\bar \sigma }_\mu }} \right|2} \right]} \right.
\end{equation}
and
\begin{equation}\label{qVqbar2}
i{M_\mu }\left( {1_q^ + ,2_{\bar q}^ - } \right) = i\left\langle {\left. {2\left| {{{\bar \sigma }_\mu }} \right|1} \right]} \right. = i\left. {\left[ {1\left| {{\sigma _\mu }} \right|2} \right.} \right\rangle.
\end{equation}
As can be seen, this scheme is flavor independent.

\subsubsection{$qg\bar q$ Blocks}
The color-ordered vertex is just the conventional vertex shown in \cite{peskin2} with the color factor and the coupling constant (in our convention) stripped. The non-vanishing blocks are
\begin{equation}
iM\left( {1_q^ - ,2_g^ - ,3_{\bar q}^ + } \right) = \frac{i}{{\sqrt 2 }}\left\langle 1 \right|{{\bar \varepsilon }_ - }\left( {2,r} \right)\left| {\left. 3 \right]} \right. =  - i\frac{{{{\left\langle {12} \right\rangle }^2}}}{{\left\langle {31} \right\rangle }},
\end{equation}
\begin{equation}
iM\left( {1_q^ + ,2_g^ + ,3_{\bar q}^ - } \right) = \frac{i}{{\sqrt 2 }}\left. {\left[ 1 \right.} \right|{\varepsilon _ + }\left( {2,r} \right)\left| 3 \right\rangle  = i\frac{{{{\left[ {12} \right]}^2}}}{{\left[ {31} \right]}},
\end{equation}
\begin{equation}
iM\left( {1_q^ - ,2_g^ + ,3_{\bar q}^ + } \right) = \frac{i}{{\sqrt 2 }}\left\langle 1 \right|{{\bar \varepsilon }_ + }\left( {2,r} \right)\left| {\left. 3 \right]} \right. =  - i\frac{{{{\left[ {23} \right]}^2}}}{{\left[ {31} \right]}}
\end{equation}
and
\begin{equation}
iM\left( {1_q^ + ,2_g^ - ,3_{\bar q}^ - } \right) = \frac{i}{{\sqrt 2 }}\left. {\left[ 1 \right.} \right|{\varepsilon _ - }\left( {2,r} \right)\left| 3 \right\rangle  = i\frac{{{{\left\langle {23} \right\rangle }^2}}}{{\left\langle {31} \right\rangle }}
\end{equation}
in which the rescaled operators $T^{a} = {\sqrt 2 }t^{a}$ were used (see Section \ref{convampl}).

\subsubsection{$ggg$ Blocks}
The color-ordered vertex in our convention is obtained from the conventional rules in \cite{peskin2} by replacing incoming by outgoing momenta and introducing particle labels growing in the clockwise order. After that, the structure constant $f^{abc}$ of the SU(3) lie algebra is replaced by the trace \cite{peskin}
\begin{equation}
- {f^{abc}} = \frac{i}{{\sqrt 2 }}Tr\left\{ {{T^a}\left[ {{T^b},{T^c}} \right]} \right\},
\end{equation}
in which, as stated before, the rescaled operators $T^{a}$ were used (see Section \ref{convampl}). Then, the color structure and the coupling constant are stripped, producing the $ggg$ blocks
\begin{equation}
\begin{split}
iM\left( {1_g^{_{{h_1}}},2_g^{_{{h_2}}},3_g^{_{{h_3}}}} \right) &=
\\
&\frac{i}{{\sqrt 2 }}{\varepsilon _{{h_1}}}\left( {{p_1}} \right) \cdot {\varepsilon _{{h_2}}}\left( {{p_2}} \right)\left( {{p_1} - {p_2}} \right) \cdot {\varepsilon _{{h_3}}}\left( {{p_3}} \right)
\\
+ &\frac{i}{{\sqrt 2 }}{\varepsilon _{{h_2}}}\left( {{p_2}} \right) \cdot {\varepsilon _{{h_3}}}\left( {{p_3}} \right)\left( {{p_2} - {p_3}} \right) \cdot {\varepsilon _{{h_1}}}\left( {{p_1}} \right)
\\
+ &\frac{i}{{\sqrt 2 }}{\varepsilon _{{h_3}}}\left( {{p_3}} \right) \cdot {\varepsilon _{{h_1}}}\left( {{p_1}} \right)\left( {{p_3} - {p_1}} \right) \cdot {\varepsilon _{{h_2}}}\left( {{p_2}} \right).
\end{split}
\end{equation}
For specific helicities it is found
\begin{equation}\label{blockggg1}
iM\left( {1_g^ - ,2_g^ - ,3_g^ + } \right) = i\frac{{{{\left\langle {12} \right\rangle }^4}}}{{\left\langle {12} \right\rangle \left\langle {23} \right\rangle \left\langle {31} \right\rangle }}
\end{equation}
and
\begin{equation}\label{blockggg2}
iM\left( {1_g^ + ,2_g^ + ,3_g^ - } \right) =  - i\frac{{{{\left[ {12} \right]}^4}}}{{\left[ {12} \right]\left[ {23} \right]\left[ {31} \right]}}.
\end{equation}
These are (up to cyclic reorderings, see Section \ref{symmetries}) the only non-vanishing blocks. As it was pointed out before, these can be seen as particular cases of the MHV amplitudes and the complex momenta is essential in their definition.

\subsection{Symmetries}\label{symmetries}

\smallskip
As can be seen, there are some symmetries that relate these basic blocks. In the spirit of reducing the number of amplitudes, mixing several symmetries is also a valuable tool.

\smallskip
For instance, parity transformation changes each helicity into its opposite. In a partial amplitude all the coupling factors that may express the lack of this symmetry are already stripped out. Then, it can be expected that a partial amplitude has the same norm that the one with opposite helicities. The specific relation between these ones is shown by the basic blocks from the previous section; observe that (treating with care the $i$ factors on both sides of the expressions) they are the complex conjugate of each other. Clearly, in more elaborated amplitudes, this conjugation (often known as parity conjugation) does not change those terms that are not affected by parity transformation and, in consequence, under parity conjugation are not replaced by their conjugates (like the complex width terms in the propagators of massive bosons). Then, in general obtaining the parity conjugate of an amplitude tends to be a simple task taking into account the behavior of spinor products under complex conjugation in the chosen convention (see (\ref{conjug1}), (\ref{conjug2}) and (\ref{conjug3})).

\smallskip
Line reversal symmetry or reflection symmetry \cite{dixon1,glover} relates a partial amplitude for the production of a set of particles in a specific order with the one for the production of the reversed order with quarks being replaced by antiquarks. This may be written as
\begin{multline}
i{M_{\mu ...\nu }}\left( {1_q^{{h_1}},2_g^{{h_2}},...,m_{\bar q}^{{h_m}},...,\left( {m + 1} \right)_q^{{h_{m + 1}}},\left( {m + 2} \right)_g^{{h_{m + 2}}},...,n_{\bar q}^{{h_n}}} \right) =
\\
{\left( { - 1} \right)^\sigma }i{M_{\nu ...\mu }}\left( {n_q^{{h_n}},...,\left( {m + 2} \right)_g^{{h_{m + 2}}},\left( {m + 1} \right)_{\bar q}^{{h_{m + 1}}},...,m_q^{{h_m}},...,2_g^{{h_2}},1_{\bar q}^{{h_1}}} \right),
\end{multline}
in which the specific phase factor ${\left( { - 1} \right)^\sigma }$ that states this relation is given by the basic blocks used in the construction of the amplitudes and depends on the kind of particles involved (also, it may be different in other spinor conventions). For instance, consider how the amplitudes for the blocks $q{V_\mu }\bar q$ or $Q{V_\mu }\bar q$ are related between each other ((\ref{qVqbar1}) and (\ref{qVqbar2})); as can be seen, the reversed order of particles is obtained with no additional phase. The phases for line reversal in QCD partial amplitudes are well known. In our convention, QCD blocks come with $(-1) ^{ n_q+ n_g -1}$ and $(-1) ^{ n_g}$ in the $qg...g\bar q$ and $gg...g$ amplitudes respectively, with $n_q$ the number of quarks and antiquarks and $n_g$ the number of gluons (see the basic QCD blocks in the previous section). These phases will provide the adequate factors in order to take into account this symmetry in more elaborated amplitudes. For instance, one of interest in the kind of processes we are treating is the associated with the production of one vector boson, a number $n_q$ of quarks and antiquarks, and a number $n_g$ of gluons. The phase that relates this amplitude with the reversed one is (in our convention) $(-1) ^{ n_q+ n_g }$ \cite{glover}.

\smallskip
As the matter of fact, line reversal is seen even in subamplitudes as will be shown in the following computations. This is a particularly useful feature of this symmetry. In fact, the blocks $q{V_\mu }\bar q$ or $Q{V_\mu }\bar q$ and $qg\bar q$ display the effect of charge conjugation transformation. For instance, consider the amplitude in which a single fermion pair coupled to a gauge boson is produced and the amplitude for the production of the reversed (in the sense stated before) fermion pair. According to the phases stated before, it can be obtained one from the other by just interchanging the corresponding fermion labels in the known amplitude; in this particular case, the reversed line is just the pair quark-antiquark coupled to the boson \cite{dixon1}.

\smallskip
We must mention also cyclic reordering. The use of this relation is particularly important when working on unpolarized cross sections, in which all the helicities configurations must be taken into account. Often, the only difference between two amplitudes is just a cyclic reordering in the set of colored particles. As can be expected, this amounts effectively to a reordering of particle labels so different amplitudes may be obtained by the adequate renaming of spinors and momenta (cyclic character obeys the structure of color ordered amplitudes). For instance, in the basic blocks for three gluons ((\ref{blockggg1}) and (\ref{blockggg2})) the remaining non-vanishing amplitudes may be obtained as cyclic reorderings of the shown amplitudes.

\subsection{Higgs Blocks}

\begin{figure}
  \centering
    \includegraphics[width=0.32\textwidth,viewport=120 45 350 330,clip]{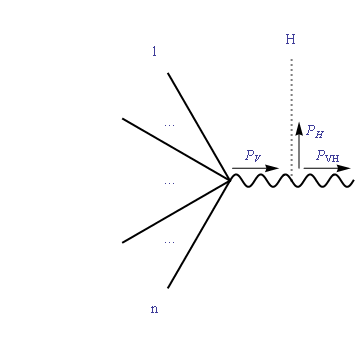}
  \caption{One possible BCFW factorization in which the particles $1,...,n$, are assumed to have no coupling with a Higgs boson, not even by an intermediate state of the sub-amplitude, except for a vector boson leg to which the scalar is coupled.}
  \label{HiggsLegs}
\end{figure}

Consider now a number of particles, $1,...,n$, that are assumed to have no coupling with a Higgs boson, as shown in Figure~\ref{HiggsLegs}. Let $i{M^\mu }\left( {p_1^{{h_1}},p_2^{{h_2}},...,p_n^{{h_n}}} \right)$ be the partial amplitude associated with the production of the particles ${p_1^{{h_1}},p_2^{{h_2}},...,p_n^{{h_n}}}$ and a vector boson ${V_\mu }$. Also, assume that any of the possible intermediate states in the last amplitude can't be coupled to the Higgs scalar (consider the cases in which the theory, the order or the kind of desired interaction forbids those couplings; in the present case the order of interest will impede them). Then, the corresponding amplitude for the production of all these particles and a Higgs boson is written then as
\begin{equation}
\begin{split}
iM_V^\tau \left( {H,p_1^{{h_1}},p_2^{{h_2}},...,p_n^{{h_n}}} \right) &= i{M^\mu }\left( {p_1^{{h_1}},p_2^{{h_2}},...,p_n^{{h_n}}} \right)\frac{{\left( { - i} \right){g_{\mu \nu }}i{g^{\nu \tau }}}}{{\left[ {P_V^2 - M_V^2 + i{\Gamma _V}{M_V}} \right]}}
\\
&= \frac{{i{M^\tau }\left( {p_1^{{h_1}},p_2^{{h_2}},...,p_n^{{h_n}}} \right)}}{{\left[ {P_V^2 - M_V^2 + i{\Gamma _V}{M_V}} \right]}}.
\end{split}
\end{equation}
This can be seen with the aid of the BCFW scheme. By hypothesis, the Higgs particle must be coupled to the vector boson leg (with  the couplings (\ref{couplingWHiggs}) and (\ref{couplingZHiggs})). Hence, when constructing the amplitude with this scheme, the BCFW diagrams contributing are the same than the Higgs-less case but with the Higgs leg inserted always in the side where the vector boson is attached. Observe that the momentum in the vector boson propagator before the Higgs coupling remains as ${P_V} =  - \sum\limits_{i = 1}^n {{p_i}} $	regardless of the shifts (neither the Higgs nor the vector boson momenta are going to be shifted) and that the momentum in the propagator of the BCFW factorization can be written in terms of the momenta of the factorized (left or right) amplitude with no vector boson (writing the momentum of the propagator in terms of momenta with no spinorial form, like massive bosons momenta, is discouraged since that term will be computed with other spinor products). In consequence, the result is the introduction of the Higgs vertex and the vector boson propagator in each diagram. Thus, the final amplitude with a Higgs leg may be factorized in these cases in the form shown before.

\begin{figure}
  \centering
    \includegraphics[width=0.48\textwidth,viewport=5 85 355 310,clip]{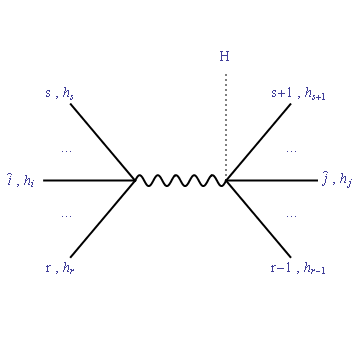}
\quad
    \includegraphics[width=0.48\textwidth,viewport=5 85 355 310,clip]{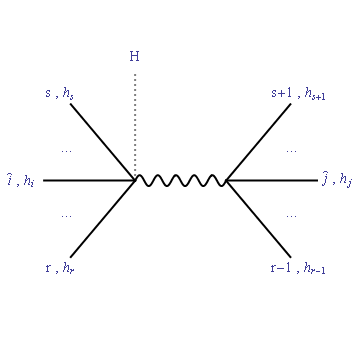}
  \caption{Two BCFW diagrams contributing to an amplitude (with the choice for $i$ and $j$ already fixed) which only differ in the position of the Higgs particle.}
  \label{Higgs2}
\end{figure}

\smallskip
There is another case of interest concerning a Higgs coupling. Consider the computation of an amplitude in which the BCFW diagrams shown in Figure~\ref{Higgs2}. These two diagrams only differ in the side in which the Higgs particle lays. As the matter of fact, this situation is not uncommon since whenever an intermediate state can be coupled to a Higgs boson, the two shown diagrams appear as different pole contributions as long as the $i$ and $j$ chosen particles lay in different sides of the factorization. Clearly, this case will also be relevant when there are legs in both sides which couple with the Higgs particle. In the present context, we'll assume that the intermediate state is a massive vector boson. If we try to write the contribution of these two diagrams we'll find
\begin{multline}
\frac{{i{M^\tau }\left( {p_r^{{h_r}},...,\hat p_{iL}^{{h_i}},...,p_s^{{h_s}}, - \hat P_L^h} \right)\left( { - i} \right)i{M_{V\tau }}\left( {H,\hat P_L^{ - h},p_{s + 1}^{{h_s}},...,\hat p_{jL}^{{h_1}},...,p_{r - 1}^{{h_{r - 1}}}} \right)}}{{\left[ {P_L^2 - M_V^2 + i{\Gamma _V}{M_V}} \right]}}
\\
+ \frac{{i{M^\tau }\left( {H,p_r^{{h_r}},...,\hat p_{iR}^{{h_i}},...,p_s^{{h_s}}, - \hat P_R^h} \right)\left( { - i} \right)i{M_{V\tau }}\left( {\hat P_R^{ - h},p_{s + 1}^{{h_s}},...,\hat p_{jR}^{{h_1}},...,p_{r - 1}^{{h_{r - 1}}}} \right)}}{{\left[ {P_R^2 - M_V^2 + i{\Gamma _V}{M_V}} \right]}}
\end{multline}
, where in each diagram we summed over the intermediate polarization states of the vector boson and assumed amplitudes which verify the Ward-Takahashi identity (as we'll see, the amplitudes found with a boson leg do vanish when contracted with the momentum of the boson particle; this is seen in a more general approach in \cite{glover}). The subindex $L$ denotes the values of the momenta in the left diagram and the subindex $R$ the values of the momenta in the right diagram. As one might expect, the shift and the intermediate momenta is different in both diagrams due to the position of the Higgs boson, despite of the fact that the $i$ and $j$ particles are the same.
With the definitions,
\begin{equation}
{P_L} = \sum\limits_{m = r}^s {{p_m}}
\end{equation}
and
\begin{equation}
{P_R} =  - \sum\limits_{m = s + 1}^{r - 1} {{p_m}}
\end{equation}
the momenta and the value of ${{z_{part}}}$ for each diagram can be written as
\begin{equation}
\hat P_{VL}^\mu  = P_R^\mu  + \frac{{{z_L}}}{2}\left\langle {\left. {i\left| {{{\bar \sigma }^\mu }} \right|j} \right]} \right.,
\end{equation}
\begin{equation}
{z_L} =  - \frac{{\left[ {P_L^2 - M_V^2 + i{\Gamma _V}{M_V}} \right]}}{{\left\langle {\left. {i\left| {{{\bar P}_L}} \right|j} \right]} \right.}},
\end{equation}
\begin{equation}
\hat P_{VR}^\mu  = P_L^\mu  + \frac{{{z_R}}}{2}\left\langle {\left. {i\left| {{{\bar \sigma }^\mu }} \right|j} \right]} \right.,
\end{equation}
\begin{equation}
{z_R} =  - \frac{{\left[ {P_R^2 - M_V^2 + i{\Gamma _V}{M_V}} \right]}}{{\left\langle {\left. {i\left| {{{\bar P}_R}} \right|j} \right]} \right.}}.
\end{equation}
Even with the generality of the situation, it is possible to obtain an interesting result. Consider the case in which the contraction
\[i{M^\tau }\left( {p_r^{{h_r}},...,\hat p_i^{{h_i}},...,p_s^{{h_s}}, - {{\hat P}^h}} \right) \times i{M_\tau }\left( {{{\hat P}^{ - h}},p_{s + 1}^{{h_s}},...,\hat p_j^{{h_1}},...,p_{r - 1}^{{h_{r - 1}}}} \right)\]
is independent of the shift in the (already fixed) $i$ and $j$ spinors (even if each amplitude depends on the shift, their contraction might not). This is not an uncommon situation as we'll see in further computations. Then, a little of algebra with the values shown before gives the natural and highly desired form for the contributions of the diagrams,
\begin{equation}
\frac{{i{M^\tau }\left( {p_r^{{h_r}},...,p_i^{{h_i}},...,p_s^{{h_s}}, - {P^h}} \right)\left( { - i} \right)i{M_\tau }\left( {{P^{ - h}},p_{s + 1}^{{h_s}},...,p_j^{{h_1}},...,p_{r - 1}^{{h_{r - 1}}}} \right)}}{{\left[ {P_L^2 - M_V^2 + i{\Gamma _V}{M_V}} \right]\left[ {P_R^2 - M_V^2 + i{\Gamma _V}{M_V}} \right]}}.
\end{equation}
Thus, the values of momenta in the vector boson propagators are the ones expected by a common Feynman amplitude. This is clearly a fundamental condition and is also an important constraint when looking for the best $(i,j)$ choice. Care must be taken when dealing with amplitudes which have massive propagators. The denominators of these last ones can't be written completely in terms of helicity spinors for obvious reasons: the mass dependent terms. Hence, except for special cases like the one treated with the Higgs particle, the algebra of this terms combined with the spinor algebra gets tedious. Then, the easiest solution is to find scenarios in which it is not necessary to operate with them, simply because the value of the momenta in the propagators is the expected one (not a possible shifted value). It is then relevant to take this into account when looking for the adequate choice for the $(i,j)$ particles that avoids shifted momenta in massive propagators.

\subsection{WBF at Tree Level}

\begin{figure}
  \centering
    \includegraphics[width=0.36\textwidth,viewport=35 100 285 290,clip]{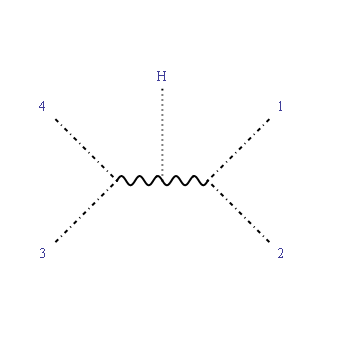}
  \caption{The only Feynman diagram which contributes to WBF at tree level (all external particles are final states). Gluons are represented with thick straight lines, fermions with dot-dashed lines, Higgs particles with dotted lines and vector bosons with wavy lines.}
  \label{FeynmanNoGluon}
\end{figure}

The effectiveness of the helicity spinor formalism is more evident as the number of external legs grows. This is particularly manifest in the kind of processes we are considering. In fact, the tree level contribution to the WBF Higgs production with no additional emissions is not obscure at all. If we focus in a specific intermediate vector boson, there is only one Feynman diagram (as stated in the conventions, all external particles are considered final states, see Figure~\ref{FeynmanNoGluon}) and the amplitude can be written directly.

\smallskip
The corresponding Feynman amplitude with the intervention of a vector boson $V$ is
\begin{multline}
i{S_V}\left( {H,1_q^{{h_1}},2_{\bar q}^{{h_2}},3_q^{{h_3}},4_{\bar q}^{{h_4}}} \right){\rm{ = }}
\\
{{\rm{e}}^3}{M_z}{H_{cf,V}}{h_{cf,V,34}}\left( { - {h_4}} \right){h_{cf,V,12}}\left( { - {h_2}} \right) i{M_V}\left( {H,1_q^{{h_1}},2_{\bar q}^{{h_2}},3_q^{{h_3}},4_{\bar q}^{{h_4}}} \right),
\end{multline}
with
\begin{multline}\label{feynampnogluon}
i{M_V}\left( {H,1_q^{{h_1}},2_{\bar q}^{{h_2}},3_q^{{h_3}},4_{\bar q}^{{h_4}}} \right) =
\\
i{\xi _V}\left( {{{P'}_V},{{P''}_V}} \right){{\bar U}^{{h_3}}}\left( {{p_3}} \right){\gamma ^\mu }{U^{ - {h_4}}}\left( {{p_4}} \right) {{\bar U}^{{h_1}}}\left( {{p_1}} \right){\gamma _\mu }{U^{ - {h_2}}}\left( {{p_2}} \right),
\end{multline}
\begin{equation}
{{P'}_V} \equiv  - {p_1} - {p_2},
\end{equation}
\begin{equation}
{{P''}_V} \equiv  - {p_3} - {p_4},
\end{equation}
and
\begin{equation}
\xi _V^{ - 1}\left( {{P_{V,1}},{P_{V,2}}} \right) \equiv \left( {P_{V,1}^2 - M_V^2 + i{\Gamma _V}{M_V}} \right)\left( {P_{V,2}^2 - M_V^2 + i{\Gamma _V}{M_V}} \right).
\end{equation}
In all the partial amplitudes, the flavor dependence is not stated explicitly, until they are inserted in the complete amplitudes. In this case, the corresponding complete amplitudes (considering all the possible intermediate bosons) can be written then as
\begin{multline}
\left( {H,{u^{{h_1}}},{{\bar d}^{{h_2}}},{d^{{h_3}}},{{\bar u}^{{h_4}}}} \right) \to
\\
i{S_W}\left( {H,{u^{{h_1}}},{{\bar d}^{{h_2}}},{d^{{h_3}}},{{\bar u}^{{h_4}}}} \right) + 2i{S_Z}\left( {H,{u^{{h_1}}},{{\bar u}^{{h_4}}},{d^{{h_3}}},{{\bar d}^{{h_2}}}} \right)
\end{multline}
and
\begin{multline}
\left( {H,u_1^{{h_1}},\bar u_2^{{h_2}},u_3^{{h_3}},\bar u_4^{{h_4}}} \right) 
\to
\\
2i{S_Z}\left( {H,u_1^{{h_1}},\bar u_2^{{h_2}},u_3^{{h_3}},\bar u_4^{{h_4}}} \right) - 2i{S_Z}\left( {H,u_1^{{h_1}},\bar u_4^{{h_4}},u_3^{{h_3}},\bar u_2^{{h_2}}} \right),
\end{multline}
where there were specified the up and down type quark relevant cases. Combinatorial factors and crossings were introduced.

\begin{figure}
  \centering
    \includegraphics[width=0.47\textwidth,viewport=0 140 360 360,clip]{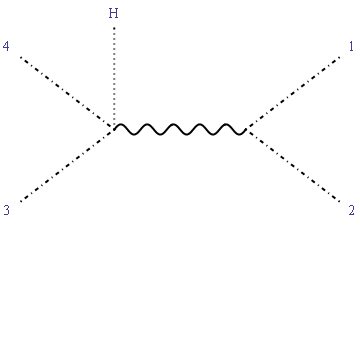}
\quad
    \includegraphics[width=0.47\textwidth,viewport=0 140 360 360,clip]{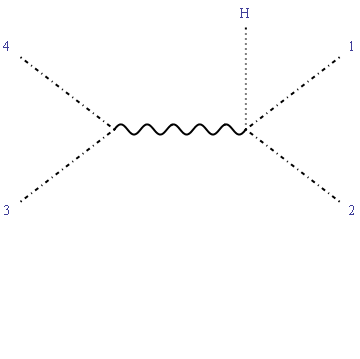}
  \caption{BCFW diagrams contributing to WBF Higgs production at tree level taking $i$ and $j$ particles as the quarks 1 and 3 (choosing the adequate order for each set of helicities and, if necessary, making a cyclic rearrangement of the diagram in order to place the $i$ particle on the left and the $j$ particle on the right. This last statement must taken into account from now on).}
  \label{BCFWNoGluon}
\end{figure}

\smallskip
What does the BCFW scheme say about this amplitude? Well, if we pick the $(i,j)$ particles as the 2 quarks, particles 1 and 3 (keeping in mind the order of helicities which guarantee the behavior in the limit as the norm of $z$ goes to infinite), there are only two diagrams, which can be seen in Figure~\ref{BCFWNoGluon}. In fact, they only differ in the position on the Higgs particle, as the case we discussed in the previous section. Furthermore, a little of algebra proves that the contractions of the amplitudes in the numerator are independent of the shift in all the possible sets of helicities. The sub-blocks come from the basic blocks we considered in the previous sections and those with a Higgs particle can be built from the prescription introduced before. Using these results, one finds the amplitude
\begin{equation}
i{M_V}\left( {H,1_q^ + ,2_{\bar q}^ - ,3_q^ + ,4_{\bar q}^ - } \right) = \frac{{ - 2i\left[ {13} \right]\left\langle {24} \right\rangle }}{{\left[ {{s_{12}} - M_V^2 + i{\Gamma _V}{M_V}} \right]\left[ {{s_{34}} - M_V^2 + i{\Gamma _V}{M_V}} \right]}}.
\end{equation}
The remaining amplitudes can be obtained with a couple of the mentioned symmetries. For instance, a fermion pair may be charge-conjugated (exchanging labels) and then the parity-conjugate of the two amplitudes (the last one and the original) complete the set of four non-vanishing amplitudes. If it is considered the line-reversed amplitude, the corresponding phase that links both amplitudes is $(-1) ^{ n_q}$ and in the following amplitudes, as gluon emissions are considered, the phases will take the form $(-1) ^{ n_q+n_g}$. All the amplitudes found were checked against the Feynman result with each set of helicities. This is done by introducing specific helicities for external particles in (\ref{feynampnogluon}).

\begin{figure}
  \centering
    \includegraphics[width=0.47\textwidth,viewport=0 140 360 360,clip]{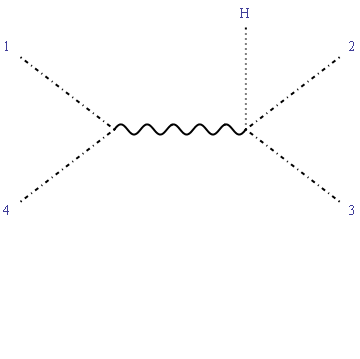}
\quad
    \includegraphics[width=0.47\textwidth,viewport=0 140 360 360,clip]{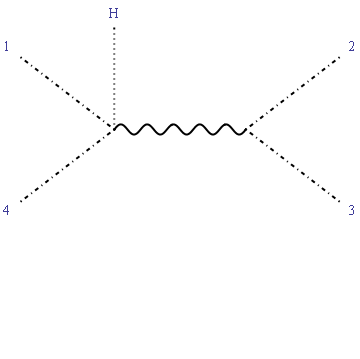}
  \caption{BCFW diagrams that do not contribute to the WBF amplitude.}
  \label{warning}
\end{figure}

\smallskip
It is relevant to point out an important warning concerning the selection of the BCFW diagrams that contribute. Observe in Figure~\ref{warning} some additional diagrams that actually hold the same order in the set of colored particles, the same order of perturbation and the same types of interactions. Notwithstanding, these last diagrams do not contribute to the amplitude we are trying to find and there are two fundamental reasons. As can be seen, these diagrams are based in crossings of particles. In first place, in the following amplitudes we'll be dealing with gluon emissions and this kind of crossings introduces a modification in the color factor dependence; the quark indices associated with the SU(3) generators would be shifted. In second place, these crossings are forbidden when dealing with a W boson; the crossed diagrams are possible only with a Z boson interaction. These contributions, as it was pointed out before, will be considered as reorderings written in terms of the basic amplitudes we are looking for with the BCFW scheme. In the following computations no further comments will be made concerning this point but it should be kept in mind.

\subsection{Gluon Emissions}

The kind of QCD tree level corrections studied were the emissions of 1 and 2 gluons.

\subsubsection{1 Gluon}

\begin{figure}
  \centering
    \includegraphics[width=0.36\textwidth,viewport=35 100 290 290,clip]{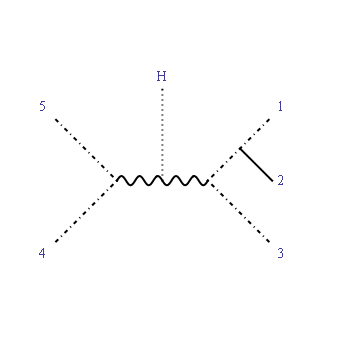}
\quad
    \includegraphics[width=0.36\textwidth,viewport=35 100 290 290,clip]{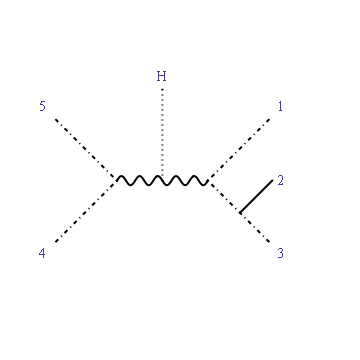}
  \caption{The Feynman diagrams that contribute to the WBF Higgs production process with one gluon emission, in which an specific color order has been considered.}
  \label{Feynman1Gluon}
\end{figure}

With one gluon emission the number of diagrams is still reduced. Since color-ordered amplitudes are related to specific color structures, it is essential to identify all the color factors involved in a particular amplitude. It is also a useful practice that leads to summarized expressions. The Feynman diagrams contributing with a specific order can be seen in Figure~\ref{Feynman1Gluon}. These provide an amplitude which is a function of the labeled particles. Then, by inserting the order of the remaining 2 Feynman diagrams (the gluon being emitted by the 4 and 5 fermion legs), it is possible to write the full amplitude in terms of these basic structures. This is the natural scheme in which the BCFW formula will provide complete amplitudes; after identifying the basic constituents (which share the same color dependence), the complete amplitude is obtained as the appropriate sum of them.

\smallskip
The amplitude according to the Feynman diagrams is
\begin{multline}
i{S_V}\left( {H,1_q^{{h_1}},2_g^ \pm ,3_{\bar q}^{{h_3}},4_q^{{h_4}},5_{\bar q}^{{h_5}}} \right){\rm{ = }}
\\
{{\rm{e}}^3}{g_{QCD}}{M_z}{H_{cf,V}}{h_{cf,V,45}}\left( { - {h_5}} \right){h_{cf,V,13}}\left( { - {h_3}} \right) T_{13}^ai{M_V}\left( {H,1_q^{{h_1}},2_g^{ \pm ,a},3_{\bar q}^{{h_3}},4_q^{{h_4}},5_{\bar q}^{{h_5}}} \right),
\end{multline}
with
\begin{multline}\label{feynamp1gluon}
i{M_V}\left( {H,1_q^{{h_1}},2_g^ \pm ,3_{\bar q}^{{h_3}},4_q^{{h_4}},5_{\bar q}^{{h_5}}} \right) =
\\
- i{\xi _V}\left( {{{P'}_V},{{P''}_V}} \right)\frac{1}{{\sqrt 2 }}{{\bar U}^{{h_4}}}\left( {{p_4}} \right){\gamma ^\mu }{U^{ - {h_5}}}\left( {{p_5}} \right) \times
\\
\times {{\bar U}^{{h_1}}}\left( {{p_1}} \right)\left\{ {{\slashed \varepsilon _ \pm }\left( {{p_2}} \right)\frac{{\left( {{\slashed p_1} + {\slashed p_2}} \right)}}{{2{p_1} \cdot {p_2}}}{\gamma _\mu } - {\gamma _\mu }\frac{{\left( {{\slashed p_2} + {\slashed p_3}} \right)}}{{2{p_2} \cdot {p_3}}}{\slashed \varepsilon _ \pm }\left( {{p_2}} \right)} \right\}{U^{ - {h_3}}}\left( {{p_3}} \right),
\end{multline}
\begin{equation}
{{P'}_V} \equiv  - {p_1} - {p_2} - {p_3},
\end{equation}
\begin{equation}
{{P''}_V} \equiv  - {p_4} - {p_5},
\end{equation}
and $\xi _V^{ - 1}\left( {{P_{V,1}},{P_{V,2}}} \right)$ as defined before. When considering all the channels, one finds the amplitudes
\begin{multline}
\left( {H,{u^{{h_1}}},{{\bar d}^{{h_3}}},{d^{{h_4}}},{{\bar u}^{{h_5}}},{g^ \pm }} \right) \to
\\
i{S_W}\left( {H,{u^{{h_1}}},{g^ \pm },{{\bar d}^{{h_3}}},{d^{{h_4}}},{{\bar u}^{{h_5}}}} \right) + i{S_W}\left( {H,{d^{{h_4}}},{g^ \pm },{{\bar u}^{{h_5}}},{u^{{h_1}}},{{\bar d}^{{h_3}}}} \right)
\\
+ 2i{S_Z}\left( {H,{u^{{h_1}}},{g^ \pm },{{\bar u}^{{h_5}}},{d^{{h_4}}},{{\bar d}^{{h_3}}}} \right) + 2i{S_Z}\left( {H,{d^{{h_4}}},{g^ \pm },{{\bar d}^{{h_3}}},{u^{{h_1}}},{{\bar u}^{{h_5}}}} \right)\
\end{multline}
and
\begin{multline}
\left( {H,u_1^{{h_1}},\bar u_3^{{h_3}},u_4^{{h_4}},\bar u_5^{{h_5}},{g^ \pm }} \right) \to
\\
2i{S_Z}\left( {H,u_1^{{h_1}},{g^ \pm },\bar u_3^{{h_3}},u_4^{{h_4}},\bar u_5^{{h_5}}} \right) + 2i{S_Z}\left( {H,u_4^{{h_4}},{g^ \pm },\bar u_5^{{h_5}},u_1^{{h_1}},\bar u_3^{{h_3}}} \right)
\\
- 2i{S_Z}\left( {H,u_1^{{h_1}},{g^ \pm },\bar u_5^{{h_5}},u_4^{{h_4}},\bar u_3^{{h_3}}} \right) - 2i{S_Z}\left( {H,u_4^{{h_4}},{g^ \pm },\bar u_3^{{h_3}},u_1^{{h_1}},\bar u_5^{{h_5}}} \right).
\end{multline}

\smallskip
With the BCFW formula, there are more possible choices for the $(i,j)$ pair in this case. For the computation of the amplitude we considered two possible approaches.

\begin{figure}
  \centering
    \includegraphics[width=0.3\textwidth,viewport=0 140 360 360,clip]{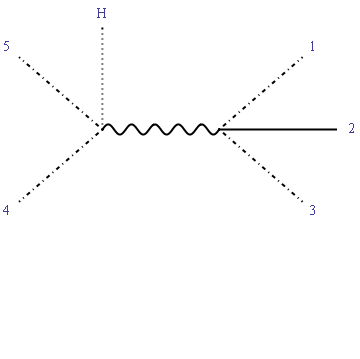}
\quad
 \includegraphics[width=0.3\textwidth,viewport=0 140 360 360,clip]{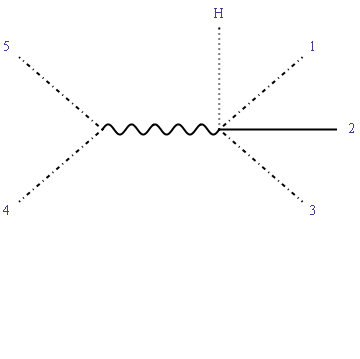}
\quad
 \includegraphics[width=0.3\textwidth,viewport=0 140 360 360,clip]{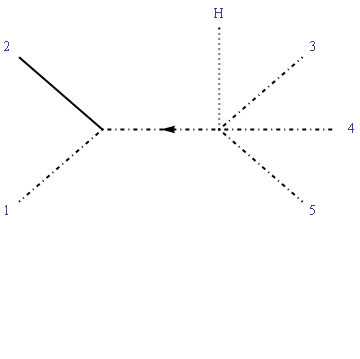}
  \caption{BCFW diagrams for the WBF Higgs production process with one gluon emission when the labeled particles are the two quarks, 1 and 4. The fermion arrow in the last diagram is relevant, as it was pointed out before.}
  \label{BCFW1Gluon_1}
\end{figure}

\smallskip
In the spirit of the WBF tree level amplitude with no gluons, one can select the particles in a way that guarantees the appearance of the BCFW diagrams with an intermediate state of a vector boson, and, therefore, of the Higgs blocks properties that we identified in the previous sections. Then one may study the possibility of choosing again the two quarks as the $i$ and $j$ particles. The BCFW diagrams then are those shown in Figure~\ref{BCFW1Gluon_1}. As can be seen, the number seems to be relatively high and, despite that some conditions facilitate their computation, the effort seems excessive.

\begin{figure}
  \centering
    \includegraphics[width=0.28\textwidth,viewport=30 95 290 260,clip]{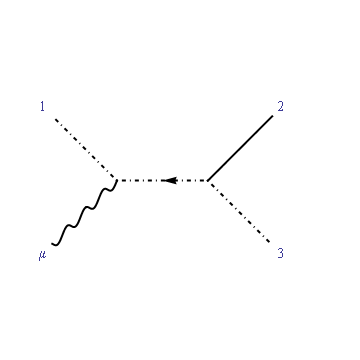}
  \caption{The BCFW diagram that contributes to the amplitude  $qg{V_\mu }\bar q$ or $Qg{V_\mu }\bar q$ at tree level.}
  \label{Glover1Gluon}
\end{figure}

\smallskip
Nevertheless, if one proceeds with this $(i,j)$ choice, there are some new blocks to be found before making the computation. The ones that were already found are the basic $qg\bar q$ and $q{V_\mu }\bar q$ or $Q{V_\mu }\bar q$ blocks, and the WBF amplitudes we found before. Those that remain are the addition of a Higgs particle to the $q{V_\mu }\bar q$ or $Q{V_\mu }\bar q$ blocks (which is done according to the prescription we introduced before) and the addition of a gluon to the same diagrams. These last ones are part of a set of amplitudes that were already found with the BCFW formula in \cite{glover}. Their construction follows the same scheme, and the diagram needed is shown in Figure~\ref{Glover1Gluon}, choosing as $i$ and $j$ particles the quark and the adjacent gluon. When they are found in our convention, they are written as

\begin{equation}
i{M_\mu }\left( {1_q^ - ,2_g^ + ,3_{\bar q}^ + } \right) =  - i\frac{{\left\langle {1\left| {{{\bar P}_V}{\sigma _\mu }} \right|1} \right\rangle }}{{\left\langle {12} \right\rangle \left\langle {23} \right\rangle }}
\end{equation}
and
\begin{equation}
i{M_\mu }\left( {1_q^ - ,2_g^ - ,3_{\bar q}^ + } \right) = i\frac{{\left[ {3\left| {{\sigma _\mu }{{\bar P}_V}} \right|3} \right]}}{{\left[ {12} \right]\left[ {23} \right]}},
\end{equation}
with ${P_V} \equiv  - {p_1} - {p_2} - {p_3}$ (the amplitudes related to the other 2 helicities sets are obtained by parity conjugation; observe that these ones are related with a composition of parity and line reversal symmetry). As can be seen, these amplitudes vanish when they are contracted with the momentum ${P_V}$ carried by the boson. With all the needed tools, the computation with this choice follows. One finds that the last diagram in Figure~\ref{BCFW1Gluon_1} vanishes and the first two can be computed with the mechanism of the previous sections.

\begin{figure}
  \centering
    \includegraphics[width=0.4\textwidth,viewport=0 140 360 355,clip]{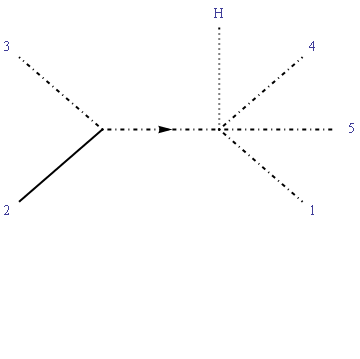}
  \caption{The only BCFW diagram that contributes to the WBF Higgs production process with one gluon emission, choosing the $i$ and $j$ particles between the gluon and the adjacent quark 1.}
  \label{BCFW1Gluon_2}
\end{figure}

\smallskip
As it was suggested, there is a shorter path and is based in a different choice for the $i$ and $j$ particles. Clearly, the formalism is more effective when the number of diagrams is reduced by choosing adjacent particles (when it is allowed). Also, in order to avoid shifted momenta in the massive propagators, the most adequate choice are the gluon and adjacent quark. The number of diagrams: only 1, which is shown in Figure~\ref{BCFW1Gluon_2}. The amplitudes are obtained readily with the blocks we already have and the prescriptions we pointed out before for equal helicities,
\begin{equation}
i{M_V}\left( {H,1_q^ - ,2_g^ - ,3_{\bar q}^ + ,4_q^ + ,5_{\bar q}^ - } \right) = \frac{{2i\left[ {34} \right]\left. {\left[ {3\left| {{{P'}_V}} \right|5} \right.} \right\rangle }}{{\left[ {12} \right]\left[ {23} \right]}}{\xi _V}\left( {{{P'}_V},{{P''}_V}} \right),
\end{equation}
\begin{equation}
i{M_V}\left( {H,1_q^ - ,2_g^ + ,3_{\bar q}^ + ,4_q^ + ,5_{\bar q}^ - } \right) = \frac{{2i\left\langle {15} \right\rangle \left. {\left[ {4\left| {{{P'}_V}} \right|1} \right.} \right\rangle }}{{\left\langle {12} \right\rangle \left\langle {23} \right\rangle }}{\xi _V}\left( {{{P'}_V},{{P''}_V}} \right).
\end{equation}
The remaining 6 amplitudes are obtained by symmetry; consider the use of charge conjugation on fermions 4 and 5 and parity. All the amplitudes found were checked against the Feynman results which are obtained by introducing specific helicities in (\ref{feynamp1gluon}), making use of the respective spinors and polarization vectors.

\subsubsection{2 Gluons}

The Feynman diagrams that contribute to the tree level amplitude of the WBF process with emission of a Higgs particle and 2 gluons have two different color structures. With the aid of the color-ordered Feynman rules and the completeness relation of the SU(3) generators these are readily identified. The corresponding diagrams associated with each color factor share specific orders in the set of colored particles. These orders, in this case, are described with the two relevant scenarios: the case in which the two gluons are adjacent and the case in which they aren't.

\begin{figure}
  \centering
    \includegraphics[width=0.3\textwidth,viewport=35 100 290 290,clip]{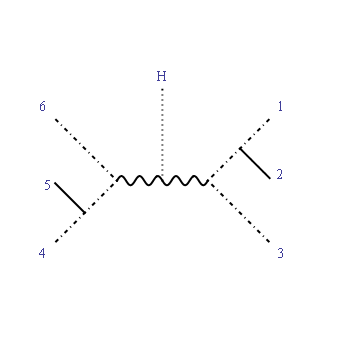}
\quad
 \includegraphics[width=0.3\textwidth,viewport=35 100 290 290,clip]{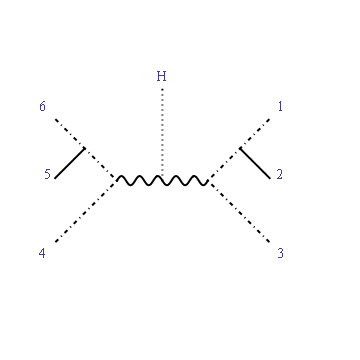}
\qquad
 \includegraphics[width=0.3\textwidth,viewport=35 100 290 290,clip]{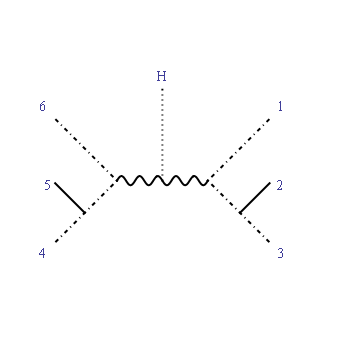}
\quad
 \includegraphics[width=0.3\textwidth,viewport=35 100 290 290,clip]{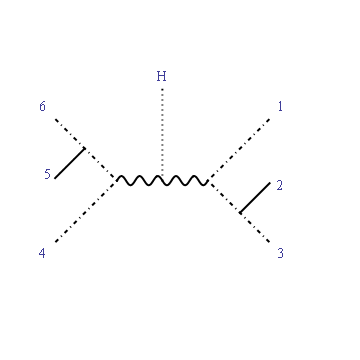}
  \caption{Feynman diagrams (with a specific color order) in the Not-Adjacent gluons amplitude of the WBF Higgs production process with two gluon emissions.}
  \label{Feynman2Gluons_1}
\end{figure}

\smallskip
In the case of 2 Not-Adjacent gluons, the Feynman diagrams with specific order that contribute are shown in Figure~\ref{Feynman2Gluons_1}. The associated amplitude is

\begin{multline}
i{S_V}\left( {H,1_q^{{h_1}},2_g^{{h_2}},3_{\bar q}^{{h_3}},4_q^{{h_4}},5_g^{{h_5}},6_{\bar q}^{{h_6}}} \right){\rm{ = }}{{\rm{e}}^3}g_{QCD}^2{M_z}{H_{cf,V}}{h_{cf,V,46}}\left( { - {h_6}} \right) \times
\\
\times {h_{cf,V,13}}\left( { - {h_3}} \right) T_{13}^aT_{46}^bi{M_V}\left( {H,1_q^{{h_1}},2_g^{{h_2},a},3_{\bar q}^{{h_3}},4_q^{{h_4}},5_g^{{h_5},b},6_{\bar q}^{{h_6}}} \right),
\end{multline}
with
\begin{multline}\label{AmpNotAdjGluonN}
i{M_V}\left( {H,1_q^{{h_1}},2_g^{{h_2}},3_{\bar q}^{{h_3}},4_q^{{h_4}},5_g^{{h_5}},6_{\bar q}^{{h_6}}} \right) = i{\xi _V}\left( {{{P'}_V},{{P''}_V}} \right)\frac{1}{2} \times
\\
\times {{\bar U}^{{h_4}}}\left( {{p_4}} \right)\left\{ {{\slashed \varepsilon _{{h_5}}}\left( {{p_5}} \right)\frac{{\left( {{\slashed p_4} + {\slashed p_5}} \right)}}{{2{p_4} \cdot {p_5}}}{\gamma ^\mu } - {\gamma ^\mu }\frac{{\left( {{\slashed p_5} + {\slashed p_6}} \right)}}{{2{p_5} \cdot {p_6}}}{\slashed \varepsilon _{{h_5}}}\left( {{p_5}} \right)} \right\}{U^{ - {h_6}}}\left( {{p_6}} \right) \times 
\\
\times {{\bar U}^{{h_1}}}\left( {{p_1}} \right)\left\{ {{\slashed \varepsilon _{{h_2}}}\left( {{p_2}} \right)\frac{{\left( {{\slashed p_1} + {\slashed p_2}} \right)}}{{2{p_1} \cdot {p_2}}}{\gamma _\mu } - {\gamma _\mu }\frac{{\left( {{\slashed p_2} + {\slashed p_3}} \right)}}{{2{p_2} \cdot {p_3}}}{\slashed \varepsilon _{{h_2}}}\left( {{p_2}} \right)} \right\}{U^{ - {h_3}}}\left( {{p_3}} \right),
\end{multline}
\begin{equation}
{{P'}_V} \equiv  - {p_1} - {p_2} - {p_3},
\end{equation}
and
\begin{equation}
{{P''}_V} \equiv  - {p_4} - {p_5} - {p_6}.
\end{equation}

\begin{figure}
  \centering
    \includegraphics[width=0.3\textwidth,viewport=40 100 290 290,clip]{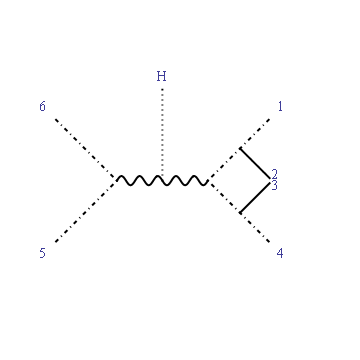}
\quad
 \includegraphics[width=0.3\textwidth,viewport=40 100 290 290,clip]{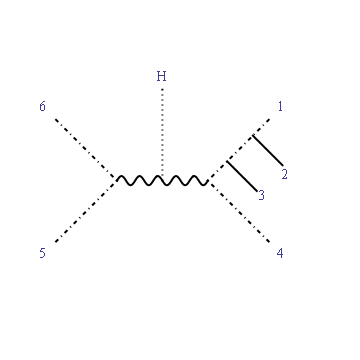}
\quad
 \includegraphics[width=0.3\textwidth,viewport=40 100 290 290,clip]{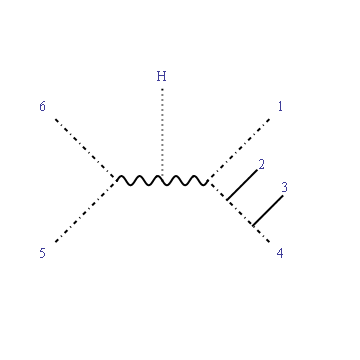}
  \caption{Feynman diagrams (with a specific color order) with no 3 gluon vertices in the Adjacent gluons amplitude of the WBF Higgs production process with two gluon emissions.}
  \label{Feynman2Gluons_2}
\end{figure}

\smallskip
In the 2 Adjacent gluon case, the Feynman diagrams involved can be divided in two groups. The first collection are the ones shown in Figure~\ref{Feynman2Gluons_2}. The color dependence is seen directly from the Feynman diagrams. The corresponding amplitude is
\begin{multline}
iS_V^{\left( I \right)}\left( {H,1_q^{{h_1}},2_g^{{h_2}},3_g^{{h_3}},4_{\bar q}^{{h_4}},5_q^{{h_5}},6_{\bar q}^{{h_6}}} \right){\rm{ = }}{{\rm{e}}^3}g_{QCD}^2{M_z}{H_{cf,V}}{h_{cf,V,56}}\left( { - {h_6}} \right) \times
\\
\times {h_{cf,V,14}}\left( { - {h_4}} \right) T_{1d}^aT_{d4}^biM_V^{\left( I \right)}\left( {H,1_q^{{h_1}},2_g^{{h_2},a},3_g^{{h_3},b},4_{\bar q}^{{h_4}},5_q^{{h_5}},6_{\bar q}^{{h_6}}} \right),
\end{multline}
with
\begin{multline}\label{feynamp2gluons1}
iM_V^{\left( I \right)}\left( {H,1_q^{{h_1}},2_g^{{h_2}},3_g^{{h_3}},4_{\bar q}^{{h_4}},5_q^{{h_5}},6_{\bar q}^{{h_6}}} \right) =
\\
i{\xi _V}\left( {{{K'}_V},{{K''}_V}} \right)\frac{1}{2}{{\bar U}^{{h_5}}}\left( {{p_5}} \right){\gamma ^\mu }{U^{ - {h_6}}}\left( {{p_6}} \right) {{\bar U}^{{h_1}}}\left( {{p_1}} \right) \times 
\\
\times \{ {\slashed \varepsilon _{{h_2}}}\left( {{p_2}} \right)\frac{{\left( {{\slashed p_1} + {\slashed p_2}} \right)}}{{2{p_1} \cdot {p_2}}}{\slashed \varepsilon _{{h_3}}}\left( {{p_3}} \right)\frac{{\left( {{\slashed p_1} + {\slashed p_2} + {\slashed p_3}} \right)}}{{2{p_1} \cdot {p_2} + 2{p_2} \cdot {p_3} + 2{p_1} \cdot {p_3}}}{\gamma _\mu }
\\
- {\slashed \varepsilon _{{h_2}}}\left( {{p_2}} \right)\frac{{\left( {{\slashed p_1} + {\slashed p_2}} \right)}}{{2{p_1} \cdot {p_2}}}{\gamma _\mu }\frac{{\left( {{\slashed p_3} + {\slashed p_4}} \right)}}{{2{p_3} \cdot {p_4}}}{\slashed \varepsilon _{{h_3}}}\left( {{p_3}} \right)
\\
+ {\gamma _\mu }\frac{{\left( {{\slashed p_2} + {\slashed p_3} + {\slashed p_4}} \right)}}{{2{p_2} \cdot {p_3} + 2{p_3} \cdot {p_4} + 2{p_2} \cdot {p_4}}}{\slashed \varepsilon _{{h_2}}}\left( {{p_2}} \right)\frac{{\left( {{\slashed p_3} + {\slashed p_4}} \right)}}{{2{p_3} \cdot {p_4}}}{\slashed \varepsilon _{{h_3}}}\left( {{p_3}} \right){U^{ - {h_4}}}\left( {{p_4}} \right)\}{U^{ - {h_4}}}\left( {{p_4}} \right),
\end{multline}
\begin{equation}
{{K'}_V} \equiv  - {p_1} - {p_2} - {p_3} - {p_4},
\end{equation}
and
\begin{equation}
{{K''}_V} \equiv  - {p_5} - {p_6}.
\end{equation}

\begin{figure}
  \centering
    \includegraphics[width=0.3\textwidth,viewport=40 100 300 290,clip]{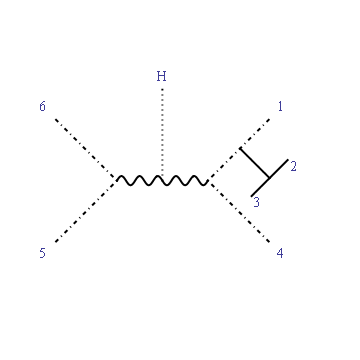}
\quad
 \includegraphics[width=0.3\textwidth,viewport=40 100 300 290,clip]{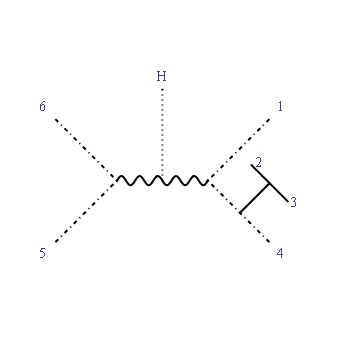}
  \caption{Feynman diagrams with 3 gluon vertices in the Adjacent gluons amplitude of the WBF Higgs production process with two gluon emissions.}
  \label{Feynman2Gluons_3}
\end{figure}

\smallskip
The second group of diagrams in the Adjacent gluons case is shown in Figure~\ref{Feynman2Gluons_3}. With the help of some of the SU(3) generators identities, the real color dependence is manifest and the corresponding amplitude can be written as the sum of two different contributions,
\begin{multline}
iS_V^{\left( {II} \right)}\left( {H,1_q^{{h_1}},2_g^{{h_2}},3_g^{{h_3}},4_{\bar q}^{{h_4}},5_q^{{h_5}},6_{\bar q}^{{h_6}}} \right) {\rm{ = }}
\\
iS_{V,ColorOrd}^{\left( {II} \right)}\left( {H,1_q^{{h_1}},2_g^{{h_2}},3_g^{{h_3}},4_{\bar q}^{{h_4}},5_q^{{h_5}},6_{\bar q}^{{h_6}}} \right)
\\
+ iS_{V,ColorOrd}^{\left( {II} \right)}\left( {H,1_q^{{h_1}},3_g^{{h_3}},2_g^{{h_2}},4_{\bar q}^{{h_4}},5_q^{{h_5}},6_{\bar q}^{{h_6}}} \right),
\end{multline}
with
\begin{multline}
iS_{V,ColorOrd}^{\left( {II} \right)}\left( {H,1_q^{{h_1}},Q_g^{{h_Q}},R_g^{{h_R}},4_{\bar q}^{{h_4}},5_q^{{h_5}},6_{\bar q}^{{h_6}}} \right) = {{\rm{e}}^3}g_{QCD}^2{M_z}{H_{cf,V}} \times
\\
\times {h_{cf,V,56}} \left( { - {h_6}} \right) {h_{cf,V,14}}\left( { - {h_4}} \right)T_{1d}^aT_{d4}^biM_V^{\left( {II} \right)}\left( {H,1_q^{{h_1}},Q_g^{{h_Q},a},R_g^{{h_R},b},4_{\bar q}^{{h_4}},5_q^{{h_5}},6_{\bar q}^{{h_6}}} \right)
\end{multline}
and
\begin{multline}
iM_V^{\left( {II} \right)}\left( {H,1_q^{{h_1}},2_g^{{h_2}},3_g^{{h_3}},4_{\bar q}^{{h_4}},5_q^{{h_5}},6_{\bar q}^{{h_6}}} \right) =
\\
- i{\xi _V}\left( {{{K'}_V},{{K''}_V}} \right)\frac{1}{2}{{\bar U}^{{h_5}}}\left( {{p_5}} \right){\gamma ^\mu }{U^{ - {h_6}}}\left( {{p_6}} \right) {{\bar U}^{{h_1}}}\left( {{p_1}} \right) \times
\\
\times \{ \{ {\varepsilon _{{h_2}}}\left( {{p_2}} \right) \cdot {\varepsilon _{{h_3}}}\left( {{p_3}} \right)\left( {{\slashed p_2} - {\slashed p_3}} \right) + {\slashed \varepsilon _{{h_3}}}\left( {{p_3}} \right)\left( {{p_2} + 2{p_3}} \right) \cdot {\varepsilon _{{h_2}}}\left( {{p_2}} \right)
\\
- {\slashed \varepsilon _{{h_2}}}\left( {{p_2}} \right)\left( {2{p_2} + {p_3}} \right) \cdot {\varepsilon _{{h_3}}}\left( {{p_3}} \right)\} \frac{{\left( {{\slashed p_1} + {\slashed p_2} + {\slashed p_3}} \right){\gamma _\mu }}}{{\left( {2{p_2} \cdot {p_3}} \right)\left( {2{p_1} \cdot {p_2} + 2{p_2} \cdot {p_3} + 2{p_1} \cdot {p_3}} \right)}}
\\
- \frac{{{\gamma _\mu }\left( {{\slashed p_2} + {\slashed p_3} + {\slashed p_4}} \right)}}{{\left( {2{p_2} \cdot {p_3} + 2{p_3} \cdot {p_4} + 2{p_2} \cdot {p_4}} \right)\left( {2{p_2} \cdot {p_3}} \right)}}\{ {\varepsilon _{{h_2}}}\left( {{p_2}} \right) \cdot {\varepsilon _{{h_3}}}\left( {{p_3}} \right)\left( {{\slashed p_2} - {\slashed p_3}} \right)
\\
+ {\slashed \varepsilon _{{h_3}}}\left( {{p_3}} \right)\left( {{p_2} + 2{p_3}} \right) \cdot {\varepsilon _{{h_2}}}\left( {{p_2}} \right) - {\slashed \varepsilon _{{h_2}}}\left( {{p_2}} \right)\left( {2{p_2} + {p_3}} \right) \cdot {\varepsilon _{{h_3}}}\left( {{p_3}} \right)\} \} {U^{ - {h_4}}}\left( {{p_4}} \right).
\end{multline}
As can be seen, there are two different color orders in this Feynman amplitude. Then it is natural to organize the amplitude according to our criteria and define (by selecting a single order) the amplitude
\begin{multline}
i{S_V}\left( {H,1_q^{{h_1}},2_g^{{h_2}},3_g^{{h_3}},4_{\bar q}^{{h_4}},5_q^{{h_5}},6_{\bar q}^{{h_6}}} \right)
\\
= iS_V^{\left( I \right)}\left( {H,1_q^{{h_1}},2_g^{{h_2}},3_g^{{h_3}},4_{\bar q}^{{h_4}},5_q^{{h_5}},6_{\bar q}^{{h_6}}} \right)
\\
+ iS_{V,ColorOrd}^{\left( {II} \right)}\left( {H,1_q^{{h_1}},2_g^{{h_2}},3_g^{{h_3}},4_{\bar q}^{{h_4}},5_q^{{h_5}},6_{\bar q}^{{h_6}}} \right),
\end{multline}
in which, clearly, the partial amplitude we are going to find with the BCFW scheme is
\begin{multline}\label{feynamp2gluons2}
i{M_V}\left( {H,1_q^{{h_1}},2_g^{{h_2}},3_g^{{h_3}},4_{\bar q}^{{h_4}},5_q^{{h_5}},6_{\bar q}^{{h_6}}} \right)
\\
= iM_V^{\left( I \right)}\left( {H,1_q^{{h_1}},2_g^{{h_2}},3_g^{{h_3}},4_{\bar q}^{{h_4}},5_q^{{h_5}},6_{\bar q}^{{h_6}}} \right)
\\
+ iM_{V,ColorOrd}^{\left( {II} \right)}\left( {H,1_q^{{h_1}},2_g^{{h_2}},3_g^{{h_3}},4_{\bar q}^{{h_4}},5_q^{{h_5}},6_{\bar q}^{{h_6}}} \right).
\end{multline}

\smallskip
The complete amplitudes with all the possible channels are then written as
\begin{multline}
\left( {H,{u^{{h_1}}},{{\bar d}^{{h_3}}},{d^{{h_4}}},{{\bar u}^{{h_6}}},g_2^{{h_2}},g_5^{{h_5}}} \right) \to 
\\
i{S_W}\left( {H,{u^{{h_1}}},g_2^{{h_2}},{{\bar d}^{{h_3}}},{d^{{h_4}}},g_5^{{h_5}},{{\bar u}^{{h_6}}}} \right) + i{S_W}\left( {H,{u^{{h_1}}},g_5^{{h_5}},{{\bar d}^{{h_3}}},{d^{{h_4}}},g_2^{{h_2}},{{\bar u}^{{h_6}}}} \right) 
\\
+ 2i{S_Z}\left( {H,{u^{{h_1}}},g_2^{{h_2}},{{\bar u}^{{h_6}}},{d^{{h_4}}},g_5^{{h_5}},{{\bar d}^{{h_3}}}} \right) + 2i{S_Z}\left( {H,{u^{{h_1}}},g_5^{{h_5}},{{\bar u}^{{h_6}}},{d^{{h_4}}},g_2^{{h_2}},{{\bar d}^{{h_3}}}} \right)
\\
+ i{S_W}\left( {H,{u^{{h_1}}},g_2^{{h_2}},g_5^{{h_5}},{{\bar d}^{{h_3}}},{d^{{h_4}}},{{\bar u}^{{h_6}}}} \right) + i{S_W}\left( {H,{u^{{h_1}}},g_5^{{h_5}},g_2^{{h_2}},{{\bar d}^{{h_3}}},{d^{{h_4}}},{{\bar u}^{{h_6}}}} \right)
\\
+ i{S_W}\left( {H,{d^{{h_4}}},g_2^{{h_2}},g_5^{{h_5}},{{\bar u}^{{h_6}}},{u^{{h_1}}},{{\bar d}^{{h_3}}}} \right) + i{S_W}\left( {H,{d^{{h_4}}},g_5^{{h_5}},g_2^{{h_2}},{{\bar u}^{{h_6}}},{u^{{h_1}}},{{\bar d}^{{h_3}}}} \right)
\\
+ 2i{S_Z}\left( {H,{u^{{h_1}}},g_2^{{h_2}},g_5^{{h_5}},{{\bar u}^{{h_6}}},{d^{{h_4}}},{{\bar d}^{{h_3}}}} \right) + 2i{S_Z}\left( {H,{u^{{h_1}}},g_5^{{h_5}},g_2^{{h_2}},{{\bar u}^{{h_6}}},{d^{{h_4}}},{{\bar d}^{{h_3}}}} \right)
\\
+ 2i{S_Z}\left( {H,{d^{{h_4}}},g_2^{{h_2}},g_5^{{h_5}},{{\bar d}^{{h_3}}},{u^{{h_1}}},{{\bar u}^{{h_6}}}} \right) + 2i{S_Z}\left( {H,{d^{{h_4}}},g_5^{{h_5}},g_2^{{h_2}},{{\bar d}^{{h_3}}},{u^{{h_1}}},{{\bar u}^{{h_6}}}} \right)
\end{multline}
and
\begin{multline}
\left( {H,u_1^{{h_1}},\bar u_3^{{h_3}},u_4^{{h_4}},\bar u_6^{{h_6}},g_2^{{h_2}},g_5^{{h_5}}} \right) \to 
\\
2i{S_Z}\left( {H,u_1^{{h_1}},g_2^{{h_2}},\bar u_3^{{h_3}},u_4^{{h_4}},g_5^{{h_5}},\bar u_6^{{h_6}}} \right) + 2i{S_Z}\left( {H,u_1^{{h_1}},g_5^{{h_5}},\bar u_3^{{h_3}},u_4^{{h_4}},g_2^{{h_2}},\bar u_6^{{h_6}}} \right)
\\
- 2i{S_Z}\left( {H,u_1^{{h_1}},g_2^{{h_2}},\bar u_6^{{h_6}},u_4^{{h_4}},g_5^{{h_5}},\bar u_3^{{h_3}}} \right) - 2i{S_Z}\left( {H,u_1^{{h_1}},g_5^{{h_5}},\bar u_6^{{h_6}},u_4^{{h_4}},g_2^{{h_2}},\bar u_3^{{h_3}}} \right)
\\
+ 2i{S_Z}\left( {H,u_1^{{h_1}},g_2^{{h_2}},g_5^{{h_5}},\bar u_3^{{h_3}},u_4^{{h_4}},\bar u_6^{{h_6}}} \right) + 2i{S_Z}\left( {H,u_1^{{h_1}},g_5^{{h_5}},g_2^{{h_2}},\bar u_3^{{h_3}},u_4^{{h_4}},\bar u_6^{{h_6}}} \right)
\\
+ 2i{S_Z}\left( {H,u_4^{{h_4}},g_2^{{h_2}},g_5^{{h_5}},\bar u_6^{{h_6}},u_1^{{h_1}},\bar u_3^{{h_3}}} \right) + 2i{S_Z}\left( {H,u_4^{{h_4}},g_5^{{h_5}},g_2^{{h_2}},\bar u_6^{{h_6}},u_1^{{h_1}},\bar u_3^{{h_3}}} \right)
\\
- 2i{S_Z}\left( {H,u_1^{{h_1}},g_2^{{h_2}},g_5^{{h_5}},\bar u_6^{{h_6}},u_4^{{h_4}},\bar u_3^{{h_3}}} \right) - 2i{S_Z}\left( {H,u_1^{{h_1}},g_5^{{h_5}},g_2^{{h_2}},\bar u_6^{{h_6}},u_4^{{h_4}},\bar u_3^{{h_3}}} \right)
\\
- 2i{S_Z}\left( {H,u_4^{{h_4}},g_2^{{h_2}},g_5^{{h_5}},\bar u_3^{{h_3}},u_1^{{h_1}},\bar u_6^{{h_6}}} \right) - 2i{S_Z}\left( {H,u_4^{{h_4}},g_5^{{h_5}},g_2^{{h_2}},\bar u_3^{{h_3}},u_1^{{h_1}},\bar u_6^{{h_6}}} \right).
\end{multline}

\begin{figure}
  \centering
    \includegraphics[width=0.4\textwidth,viewport=0 110 360 360,clip]{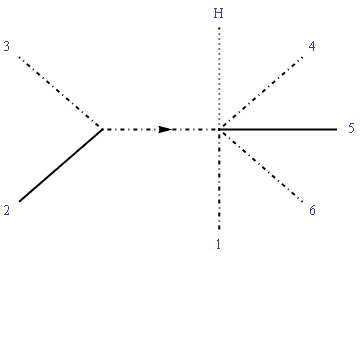}
  \caption{BCFW diagram in the Not-Adjacent gluons amplitude of the WBF Higgs production process with two gluon emissions.}
  \label{BCFW2Gluons_1}
\end{figure}

\smallskip
Now, lets find with the BCFW scheme the corresponding amplitudes. In the 2 Not-Adjacent gluons case, the obvious choice are, again, one gluon and the adjacent quark. The only diagram that contributes is shown in Figure~\ref{BCFW2Gluons_1}. The amplitudes, with ${{\mathord{\buildrel{\lower3pt\hbox{$\scriptscriptstyle\frown$}} 
\over P} }_V} \equiv {{P''}_{V,\mu }}{{\bar \sigma }^\mu }$ are

\begin{equation}\label{sym1}
i{M_V}\left( {H,1_q^ - ,2_g^ + ,3_{\bar q}^ + ,4_q^ + ,5_g^ - ,6_{\bar q}^ - } \right) = \frac{{ - 2i\left. {\left[ {4\left| {{{P'}_V}} \right|1} \right.} \right\rangle \left. {\left[ {4\left| {{{P''}_V}} \right|1} \right.} \right\rangle }}{{\left\langle {12} \right\rangle \left\langle {23} \right\rangle \left[ {45} \right]\left[ {56} \right]}}{\xi _V}\left( {{{P'}_V},{{P''}_V}} \right),
\end{equation}
\begin{equation}\label{sym2}
i{M_V}\left( {H,1_q^ + ,2_g^ + ,3_{\bar q}^ - ,4_q^ - ,5_g^ - ,6_{\bar q}^ + } \right) = \frac{{ - 2i\left. {\left[ {6\left| {{{P'}_V}} \right|3} \right.} \right\rangle \left. {\left[ {6\left| {{{P''}_V}} \right|3} \right.} \right\rangle }}{{\left\langle {12} \right\rangle \left\langle {23} \right\rangle \left[ {45} \right]\left[ {56} \right]}}{\xi _V}\left( {{{P'}_V},{{P''}_V}} \right),
\end{equation}
\begin{equation}\label{sym3}
i{M_V}\left( {H,1_q^ - ,2_g^ + ,3_{\bar q}^ + ,4_q^ - ,5_g^ - ,6_{\bar q}^ + } \right) = \frac{{2i\left. {\left[ {6\left| {{{P'}_V}} \right|1} \right.} \right\rangle \left. {\left[ {6\left| {{{P''}_V}} \right|1} \right.} \right\rangle }}{{\left\langle {12} \right\rangle \left\langle {23} \right\rangle \left[ {45} \right]\left[ {56} \right]}}{\xi _V}\left( {{{P'}_V},{{P''}_V}} \right),
\end{equation}
\begin{equation}\label{sym4}
i{M_V}\left( {H,1_q^ - ,2_g^ - ,3_{\bar q}^ + ,4_q^ - ,5_g^ - ,6_{\bar q}^ + } \right) = \frac{{2i\left[ {36} \right]\left[ {3\left| {{{P'}_V}{{\mathord{\buildrel{\lower3pt\hbox{$\scriptscriptstyle\frown$}} 
\over P} }_V}} \right|6} \right]}}{{\left[ {12} \right]\left[ {23} \right]\left[ {45} \right]\left[ {56} \right]}}{\xi _V}\left( {{{P'}_V},{{P''}_V}} \right),
\end{equation}
\begin{equation}\label{sym5}
i{M_V}\left( {H,1_q^ - ,2_g^ + ,3_{\bar q}^ + ,4_q^ - ,5_g^ + ,6_{\bar q}^ + } \right) = \frac{{2i\left\langle {14} \right\rangle \left\langle {1\left| {{{P'}_V}{{\mathord{\buildrel{\lower3pt\hbox{$\scriptscriptstyle\frown$}} 
\over P} }_V}} \right|4} \right\rangle }}{{\left\langle {12} \right\rangle \left\langle {23} \right\rangle \left\langle {45} \right\rangle \left\langle {56} \right\rangle }}{\xi _V}\left( {{{P'}_V},{{P''}_V}} \right)
\end{equation}
and
\begin{equation}\label{sym6}
i{M_V}\left( {H,1_q^ - ,2_g^ - ,3_{\bar q}^ + ,4_q^ + ,5_g^ - ,6_{\bar q}^ - } \right) = \frac{{ - 2i\left[ {34} \right]\left[ {3\left| {{{P'}_V}{{\mathord{\buildrel{\lower3pt\hbox{$\scriptscriptstyle\frown$}} 
\over P} }_V}} \right|4} \right]}}{{\left[ {12} \right]\left[ {23} \right]\left[ {45} \right]\left[ {56} \right]}}{\xi _V}\left( {{{P'}_V},{{P''}_V}} \right).
\end{equation}
The remaining 10 amplitudes can be obtained by symmetry with the use of parity and cyclic reordering. As can be seen, the number of calculated amplitudes can be reduced even more with the use compositions of line reversal, parity and cyclic reordering (like the first and second amplitudes, equations (\ref{sym1}) and (\ref{sym2}), and the fourth and fifth, equations (\ref{sym4}) and (\ref{sym5})). Observe that the pairs of amplitudes (\ref{sym1}), (\ref{sym3}) and (\ref{sym4}), (\ref{sym6}) show a form of line reversal symmetry in subamplitudes in the fermion line $4_q ,5_g ,6_{\bar q} $, with the phase expected for the amplitude of one external vector boson and QCD additional external particles, $(-1) ^{ n_q+ n_g }$, as pointed out in previous sections. All the amplitudes found were checked against the Feynman results which are obtained by introducing specific helicities in (\ref{AmpNotAdjGluonN}).

\begin{figure}
  \centering
    \includegraphics[width=0.4\textwidth,viewport=0 110 360 360,clip]{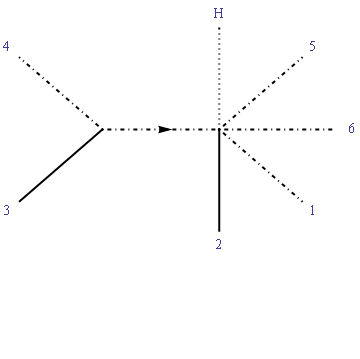}
\quad
    \includegraphics[width=0.4\textwidth,viewport=0 110 360 360,clip]{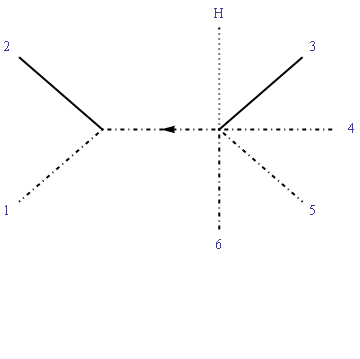}
  \caption{BCFW diagrams in the Adjacent gluons amplitude of the WBF Higgs production process with two gluon emissions.}
  \label{BCFW2Gluons_2}
\end{figure}

\smallskip
In the case of 2 Adjacent gluons, the 2 diagrams that contribute are shown in Figure~\ref{BCFW2Gluons_2}, choosing the adjacent gluons as the labeled particles (one gluon and an adjacent quark provide the same number of diagrams). The amplitudes are
\begin{equation}\label{sym21}
i{M_V}\left( {H,1_q^ - ,2_g^ - ,3_g^ - ,4_{\bar q}^ + ,5_q^ - ,6_{\bar q}^ + } \right) = \frac{{ - 2i\left[ {46} \right]\left. {\left[ {4\left| {{{K'}_V}} \right|5} \right.} \right\rangle }}{{\left[ {12} \right]\left[ {23} \right]\left[ {34} \right]}}{\xi _V}\left( {{{K'}_V},{{K''}_V}} \right),
\end{equation}
\begin{multline}
i{M_V}\left( {H,1_q^ - ,2_g^ - ,3_g^ + ,4_{\bar q}^ + ,5_q^ - ,6_{\bar q}^ + } \right) =
\\
- \frac{{2i\left[ {13} \right]\left[ {6\left| {{{K'}_V}\left( {{{\bar p}_1} + {{\bar p}_2}} \right)} \right|3} \right]\left. {\left[ {3\left| {\left( {{p_1} + {p_2}} \right)} \right|5} \right.} \right\rangle }}{{\left[ {12} \right]\left[ {23} \right]{s_{123}}\left. {\left[ {1\left| {\left( {{p_2} + {p_3}} \right)} \right|4} \right.} \right\rangle }}{\xi _V}\left( {{{K'}_V},{{K''}_V}} \right)
\\
- \frac{{2i\left\langle {24} \right\rangle \left\langle {2\left| {\left( {{{\bar p}_3} + {{\bar p}_4}} \right){{K'}_V}} \right|5} \right\rangle \left\langle {\left. {2\left| {\left( {{{\bar p}_3} + {{\bar p}_4}} \right)} \right|6} \right]} \right.}}{{\left\langle {23} \right\rangle \left\langle {34} \right\rangle {s_{234}}\left. {\left[ {1\left| {\left( {{p_2} + {p_3}} \right)} \right|4} \right.} \right\rangle }}{\xi _V}\left( {{{K'}_V},{{K''}_V}} \right),
\end{multline}
\begin{multline}
i{M_V}\left( {H,1_q^ - ,2_g^ + ,3_g^ - ,4_{\bar q}^ + ,5_q^ - ,6_{\bar q}^ + } \right) =
\\
- \frac{{2i{{\left\langle {13} \right\rangle }^3}\left[ {46} \right]\left. {\left[ {4\left| {{{K'}_V}} \right|5} \right.} \right\rangle }}{{\left\langle {12} \right\rangle \left\langle {23} \right\rangle {s_{123}}\left\langle {\left. {1\left| {\left( {{{\bar p}_2} + {{\bar p}_3}} \right)} \right|4} \right]} \right.}}{\xi _V}\left( {{{K'}_V},{{K''}_V}} \right)
\\
+ \frac{{2i{{\left[ {24} \right]}^3}\left\langle {15} \right\rangle \left. {\left[ {6\left| {{{K'}_V}} \right|1} \right.} \right\rangle }}{{\left[ {23} \right]\left[ {34} \right]{s_{234}}\left\langle {\left. {1\left| {\left( {{{\bar p}_2} + {{\bar p}_3}} \right)} \right|4} \right]} \right.}}{\xi _V}\left( {{{K'}_V},{{K''}_V}} \right)
\end{multline}
and
\begin{equation}\label{sym22}
i{M_V}\left( {H,1_q^ - ,2_g^ + ,3_g^ + ,4_{\bar q}^ + ,5_q^ - ,6_{\bar q}^ + } \right) = \frac{{2i\left\langle {15} \right\rangle \left. {\left[ {6\left| {{{K'}_V}} \right|1} \right.} \right\rangle }}{{\left\langle {12} \right\rangle \left\langle {23} \right\rangle \left\langle {34} \right\rangle }}{\xi _V}\left( {{{K'}_V},{{K''}_V}} \right).
\end{equation}
The remaining 12 amplitudes can be obtained by symmetry with the use of parity and charge conjugation between fermions 5 and 6. Observe that with the appropriate composition of parity, line reversal and cyclic reordering over the amplitude (\ref{sym21}), it can be obtained the amplitude (\ref{sym22}). All the amplitudes found were checked against the Feynman results which are obtained by introducing specific helicities in (\ref{feynamp2gluons2}).

\section{Conclusions}\label{conclusions}

We worked out a specific scattering process with the helicity spinor formalism and the BCFW recursion scheme at tree level. With a small number of BCFW diagrams (one or two), compact color-ordered amplitudes were easily obtained using the most simple blocks. As an original approach for this process, the scheme proved to be increasingly efficient as the number of external legs grows, overcoming in simplicity the conventional Feynman approach and reducing the algebra substantially. Moreover, the specific relations between amplitudes given by the symmetries were observed and characterized.

\smallskip
Currents and amplitudes were obtained with some degree of generalization and represent a step in further calculations. As one of the characteristics of the BCFW scheme, each amplitude is a step closer to more complex amplitudes. Successive  analytical corrections can be obtained with little effort and WBF process is particularly fit for this treatment. In fact, there are advanced programs based in this principles, deriving all the benefits that its effectiveness provide as in \cite{mastrolia}. 

\smallskip
The sum over intermediate states was performed over polarizations in massive particles with positive results. As the matter of fact, its use was put into practice when the choice for the labeled particles was not the most efficient, in the present context. However, it provided a quite general result concerning amplitudes of the WBF kind, and there is no reason to expect less than a major potential in this approach. Lots of topics in this area are still an object of study, such as the soft limit behavior of this spinor amplitudes \cite{boucher}, which are bringing new strategies in phenomenology. Also, processes of interest like QCD corrections to WBF with photon emissions \cite{zeppenfeld2011} are potential scenarios in which BCFW scheme might bring new and compelling results.

\appendix

\section{Some Helicity Spinors Properties and Results}\label{results}
Here we present some known results in helicity spinor formalism with the convention and definitions presented in Section \ref{conventions}.
\subsection{Weyl Spinors}\label{resultsweyl}
In first place, consider some properties of spinor products. The behavior under complex conjugation is readily understood,
\begin{equation}\label{conjug1}
{\left. {\left[ {p\left| {{\sigma ^\alpha }\left( {{{\bar \sigma }^\beta }{\sigma ^\gamma }} \right)\left( {{{\bar \sigma }^\delta }{\sigma ^\varepsilon }} \right) \cdots \left( {{{\bar \sigma }^\rho }{\sigma ^\sigma }} \right)} \right|k} \right.} \right\rangle ^*} = \left. {\left[ {k\left| {\left( {{\sigma ^\sigma }{{\bar \sigma }^\rho }} \right) \cdots \left( {{\sigma ^\varepsilon }{{\bar \sigma }^\delta }} \right)\left( {{\sigma ^\gamma }{{\bar \sigma }^\beta }} \right){\sigma ^\alpha }} \right|p} \right.} \right\rangle,
\end{equation}
\begin{equation}\label{conjug2}
{\left\langle {p\left| {\left( {{{\bar \sigma }^\beta }{\sigma ^\gamma }} \right)\left( {{{\bar \sigma }^\delta }{\sigma ^\varepsilon }} \right) \cdots \left( {{{\bar \sigma }^\rho }{\sigma ^\sigma }} \right)} \right|k} \right\rangle ^*} = \left[ {k\left| {\left( {{\sigma ^\sigma }{{\bar \sigma }^\rho }} \right) \cdots \left( {{\sigma ^\varepsilon }{{\bar \sigma }^\delta }} \right)\left( {{\sigma ^\gamma }{{\bar \sigma }^\beta }} \right)} \right|p} \right],
\end{equation}
\begin{equation}\label{conjug3}
{\left[ {p\left| {\left( {{\sigma ^\beta }{{\bar \sigma }^\gamma }} \right)\left( {{\sigma ^\delta }{{\bar \sigma }^\varepsilon }} \right) \cdots \left( {{\sigma ^\rho }{{\bar \sigma }^\sigma }} \right)} \right|k} \right]^*} = \left\langle {k\left| {\left( {{{\bar \sigma }^\sigma }{\sigma ^\rho }} \right) \cdots \left( {{{\bar \sigma }^\varepsilon }{\sigma ^\delta }} \right)\left( {{{\bar \sigma }^\gamma }{\sigma ^\beta }} \right)} \right|p} \right\rangle.
\end{equation}
In particular,
\begin{equation}
{\left\langle {pk} \right\rangle ^*} = {\left[ {u_L^\dag \left( p \right){u_R}\left( k \right)} \right]^*} = u_R^\dag \left( k \right){u_L}\left( p \right) = \left[ {kp} \right],
\end{equation}
\begin{equation}
{\left. {\left[ {p\left| {{\sigma ^\mu }} \right|k} \right.} \right\rangle ^*} = ¡u_R^\dag \left( k \right){\sigma ^{\mu \dag }}{u_R}\left( p \right) = u_R^\dag \left( k \right){\sigma ^\mu }{u_R}\left( p \right) = \left. {\left[ {k\left| {{\sigma ^\mu }} \right|p} \right.} \right\rangle.
\end{equation}

\smallskip
By projecting onto definite helicities the sum over polarization states,
\begin{equation}
{{U_R}\left( p \right){{\bar U}_R}\left( p \right) + {U_L}\left( p \right){{\bar U}_L}\left( p \right) = \slashed p},
\end{equation}
one can make the identification
\begin{equation}
\left| j \right\rangle \left. {\left[ j \right.} \right| = {p_j}_\mu {{\bar \sigma }^\mu } \equiv {{\bar p}_j},
\end{equation}
\begin{equation}
\left. {\left| j \right.} \right]\left\langle {\left. j \right|} \right. = {p_j}_\mu {\sigma ^\mu } \equiv {p_j}.
\end{equation}

\smallskip
Gordon's Identity then follows
\begin{equation}
\left\langle {\left. {j\left| {{{\bar \sigma }^\mu }} \right|j} \right]} \right. = Tr\left\{ {{{\bar \sigma }^\mu }\left. {\left| j \right.} \right]\left\langle {\left. j \right|} \right.} \right\} = {p_{j\nu }}Tr\left\{ {{{\bar \sigma }^\mu }{\sigma ^\nu }} \right\} = 2{g^{\mu \nu }}{p_{j\nu }} = 2p_j^\mu,
\end{equation}
as well as the following identity
\begin{equation}\label{sij}
{s_{ji}} \equiv \left\langle {ji} \right\rangle \left[ {ij} \right] = {p_{i\mu }}\left\langle {\left. {j\left| {{{\bar \sigma }^\mu }} \right|j} \right]} \right. =  + 2{p_j} \cdot {p_i},
\end{equation}
describing a $\frac{1}{2}$ helicity Weyl spinor $u_L \left( p \right)$ as the square root of the spin 1 four vector $p_\mu$. The phase in (\ref{sij}) agrees with QCD literature, as in \cite{dixon1,dixon2,peskin}.

\smallskip
Multiple application of $\varepsilon $ according to (\ref{leftright}) gives
\begin{equation}
\left. {\left[ {p\left| {{\sigma ^\alpha }\left( {{{\bar \sigma }^\beta }{\sigma ^\gamma }} \right)\left( {{{\bar \sigma }^\delta }{\sigma ^\varepsilon }} \right) \cdots \left( {{{\bar \sigma }^\rho }{\sigma ^\sigma }} \right)} \right|k} \right.} \right\rangle  = \left\langle {\left. {k\left| {\left( {{{\bar \sigma }^\sigma }{\sigma ^\rho }} \right) \cdots \left( {{{\bar \sigma }^\varepsilon }{\sigma ^\delta }} \right)\left( {{{\bar \sigma }^\gamma }{\sigma ^\beta }} \right){{\bar \sigma }^\alpha }} \right|p} \right]} \right.,
\end{equation}
\begin{equation}
\left\langle {p\left| {\left( {{{\bar \sigma }^\beta }{\sigma ^\gamma }} \right)\left( {{{\bar \sigma }^\delta }{\sigma ^\varepsilon }} \right) \cdots \left( {{{\bar \sigma }^\rho }{\sigma ^\sigma }} \right)} \right|k} \right\rangle  =  - \left\langle {k\left| {\left( {{{\bar \sigma }^\sigma }{\sigma ^\rho }} \right) \cdots \left( {{{\bar \sigma }^\varepsilon }{\sigma ^\delta }} \right)\left( {{{\bar \sigma }^\gamma }{\sigma ^\beta }} \right)} \right|p} \right\rangle,
\end{equation}
\begin{equation}
\left[ {p\left| {\left( {{\sigma ^\beta }{{\bar \sigma }^\gamma }} \right)\left( {{\sigma ^\delta }{{\bar \sigma }^\varepsilon }} \right) \cdots \left( {{\sigma ^\rho }{{\bar \sigma }^\sigma }} \right)} \right|k} \right] =  - \left[ {k\left| {\left( {{\sigma ^\sigma }{{\bar \sigma }^\rho }} \right) \cdots \left( {{\sigma ^\varepsilon }{{\bar \sigma }^\delta }} \right)\left( {{\sigma ^\gamma }{{\bar \sigma }^\beta }} \right)} \right|p} \right].
\end{equation}
In particular, one finds antisymmetry in
$\left\langle {pk} \right\rangle  =  - \left\langle {kp} \right\rangle $ and $\left[ {pk} \right] =  - \left[ {kp} \right]$, and the identities
\begin{equation}
\left. {\left[ {j\left| {{\sigma _\mu }} \right|k} \right.} \right\rangle  = \left\langle {\left. {k\left| {{{\bar \sigma }^\mu }} \right|j} \right]} \right.,
\end{equation}
\begin{equation}
\left\langle {j\left| {{{\bar \sigma }_\mu }{\sigma _\nu }} \right|k} \right\rangle  =  - \left\langle {k\left| {{{\bar \sigma }_\nu }{\sigma _\mu }} \right|j} \right\rangle,
\end{equation}
\begin{equation}
\left[ {j\left| {{\sigma _\mu }{{\bar \sigma }_\nu }} \right|k} \right] =  - \left[ {k\left| {{\sigma _\nu }{{\bar \sigma }_\mu }} \right|j} \right].
\end{equation}
As a consequence, one may point out a particular characteristic of linear combinations like $\left| a \right\rangle  \equiv \left| b \right\rangle  + d\left| c \right\rangle $. Contracting with an arbitrary spinor $\left| e \right\rangle $ one finds $\left\langle {ea} \right\rangle  =  - \left\{ {\left\langle b \right| + d\left\langle c \right|} \right\}\left| e \right\rangle  =  - \left\langle {ae} \right\rangle $, so $\left\langle a \right| = \left\langle b \right| + d\left\langle c \right|$, and in general $\left\langle a \right| \ne \left\langle b \right| + {d^*}\left\langle c \right|$. Hence, this angle and square notation is not related with the standard Quantum Mechanics Dirac formalism.

\smallskip
There are some well known results concerning the helicity spinors formalism. In first place, 4-component spinors verify Fierz identity
\begin{equation}
{\bar U_L}\left( k \right){\gamma ^\mu }{U_L}\left( p \right){\gamma _\mu } = 2\left[ {{U_L}\left( p \right){{\bar U}_L}\left( k \right) + {U_R}\left( k \right){{\bar U}_R}\left( p \right)} \right],
\end{equation}
which can be proved by noticing that the right side of the equation (a 4$\times$4 matrix $M$) anti-commutes with ${{\gamma ^5}}$, and therefore it can be written as \cite{peskin2} $M = {A^\mu }{\gamma _\mu } + {B^\mu }{\gamma ^5}{\gamma _\mu }$, or, with the redefinitions, ${A^\mu } = \frac{1}{2}\left( {{W^\mu } + {V^\mu }} \right)$ and ${B^\mu } = \frac{1}{2}\left( {{W^\mu } - {V^\mu }} \right)$, as $M = \frac{1}{2}\left( {1 - {\gamma ^5}} \right){\gamma _\mu }{V^\mu } + \frac{1}{2}\left( {1 + {\gamma ^5}} \right){\gamma _\mu }{W^\mu }$. By multiplying by the adequate factors and taking the trace, it is found ${V^\mu } = \frac{1}{2}Tr\left[ {{\gamma ^\mu }\frac{1}{2}\left( {1 - {\gamma ^5}} \right)M} \right]$ and ${W^\mu } = \frac{1}{2}Tr\left[ {{\gamma ^\mu }\frac{1}{2}\left( {1 + {\gamma ^5}} \right)M} \right]$. Introducing the specific matrix $M$, it is found ${V^\mu } = {{\bar U}_L}\left( k \right){\gamma ^\mu }{U_L}\left( p \right) = {{\bar U}_R}\left( p \right){\gamma ^\mu }{U_R}\left( k \right) = {W^\mu }$. In 2-component spinors this result looks like
$\left\langle {\left. {k\left| {{{\bar \sigma }^\mu }} \right|p} \right]} \right.{\gamma _\mu } = 2\left\{ {\left| {\left. p \right]} \right.\left\langle k \right| \otimes \left| k \right\rangle \left. {\left[ p \right.} \right|} \right\}$, or the notation may be relaxed by writing just
\begin{equation}
\left\langle {\left. {k\left| {{{\bar \sigma }^\mu }} \right|p} \right]} \right.{\gamma _\mu } = 2\left\{ {\left| {\left. p \right]} \right.\left\langle k \right| + \left| k \right\rangle \left. {\left[ p \right.} \right|} \right\}.
\end{equation}
In particular,
\begin{equation}
\left\langle {\left. {k\left| {{{\bar \sigma }^\mu }} \right|p} \right]} \right.\left\langle {\left. {l\left| {{{\bar \sigma }^\mu }} \right|m} \right]} \right. = 2\left\langle {lk} \right\rangle \left[ {pm} \right].
\end{equation}

\smallskip
Schouten identity is also a useful tool,
\begin{equation}
\left\langle {ij} \right\rangle \left\langle {kl} \right\rangle  + \left\langle {il} \right\rangle \left\langle {jk} \right\rangle  + \left\langle {ik} \right\rangle \left\langle {lj} \right\rangle  = 0,
\end{equation}
which is a consequence of the antisymmetry of the expression in $j$, $k$ and $l$ \cite{peskin}. It can be shown also by proving that
\begin{equation}
\left\langle {ij} \right\rangle \left[ {jk} \right]\left\langle {kl} \right\rangle \left[ {li} \right] =  - \left\langle {ik} \right\rangle \left[ {kj} \right]\left\langle {jl} \right\rangle \left[ {li} \right] + \left\langle {il} \right\rangle \left[ {li} \right]\left\langle {kj} \right\rangle \left[ {jk} \right],
\end{equation}
taking into account that trace technology allows us to write \cite{dixon1}
\begin{equation}
\left\langle {ij} \right\rangle \left[ {jk} \right]\left\langle {kl} \right\rangle \left[ {li} \right] = \frac{1}{2}\left[ {{s_{jk}}{s_{li}} - {s_{jl}}{s_{ki}} + {s_{ji}}{s_{kl}} + 4i{{\left( {{p_j}} \right)}_\mu }{{\left( {{p_k}} \right)}_\nu }{{\left( {{p_l}} \right)}_\rho }{{\left( {{p_i}} \right)}_\sigma }{\varepsilon ^{\mu \nu \rho \sigma }}} \right].
\end{equation}

\smallskip
Finally, momentum conservation is often a source of relations between spinor products. For instance, when the momenta of $n$ Weyl particles satisfy $\sum\limits_{i = 1}^n {k_i^\mu }  = 0$, it is clear that $\sum\limits_{i = 1,i \ne j,m}^n {\left[ {ji} \right]\left\langle {im} \right\rangle }  = 0$. Even more, conservation and the square of some linear combinations of momenta produce relations in terms of the ${s_{jk}}$.   

\subsection{Polarization Vectors}\label{resultspolvec}

The definitions (\ref{polr}) and (\ref{poll}) guarantee the demanded properties for polarization vectors. In first place, under conjugation
\begin{equation}
{\left[ {\varepsilon _ + ^{*\mu }\left( {k,r} \right)} \right]^*} = \varepsilon _ - ^{*\mu }\left( {k,r} \right),
\end{equation}
and the required transversality for any $r$,
\begin{equation}
\varepsilon _h^*\left( {k,r} \right) \cdot k = \varepsilon _h^{*\mu }\left( {k,r} \right){k_\mu } = 0,
\end {equation}
with the bonus
\begin{equation}
\varepsilon _h^*\left( {k,r} \right) \cdot r = 0,
\end{equation}
which is one of the benefits in the process of calculating amplitudes since many terms will vanish with the right reference choice. This is seen in the kind of contractions expected in QCD amplitudes like
\begin{equation}
\varepsilon _{{h_k}}^*\left( {k,{r_k}} \right) \cdot \varepsilon _{{h_p}}^*\left( {p,{r_p}} \right) = \left\{ \begin{array}{l}
\frac{{\left\langle {{r_k}{r_p}} \right\rangle \left[ {pk} \right]}}{{\left\langle {{r_k}k} \right\rangle \left\langle {{r_p}p} \right\rangle }},{h_k} = {h_p} =  + \\
\frac{{\left\langle {kp} \right\rangle \left[ {{r_p}{r_k}} \right]}}{{\left[ {{r_k}k} \right]\left[ {{r_p}p} \right]}},{h_k} = {h_p} =  - \\
 - \frac{{\left\langle {p{r_k}} \right\rangle \left[ {{r_p}k} \right]}}{{\left\langle {k{r_k}} \right\rangle \left[ {{r_p}p} \right]}},{h_k} =  - {h_p} =  + 
\end{array} \right.,
\end{equation}
in which antisymmetry vanishes spinor products with an adequate reference choice. 

\smallskip
Computation of gauge-invariant scattering amplitudes is not affected by the reference vector $r$ due to the Ward-Takahashi identity \cite{weinberg}. In fact,
\begin{equation}
\varepsilon _ + ^{*\mu }\left( {k,{r_1}} \right) - \varepsilon _ + ^{*\mu }\left( {k,{r_2}} \right) =  - \frac{{\sqrt 2 \left\langle {{r_1}{r_2}} \right\rangle }}{{\left\langle {{r_1}k} \right\rangle \left\langle {{r_2}k} \right\rangle }}{k^\mu },
\end{equation}
which shows that changing the reference momentum does amount to an on-shell gauge transformation. This freedom supports the effectiveness of the formalism stated before as computations are simplified with the appropriate choice of reference momenta.

\smallskip
Moreover, the behavior under rotations is the expected one. Using the fact that
\begin{equation}
{\left[ {D\left( {\phi ,\theta ,\gamma } \right)} \right]^{ - 1}}{\sigma ^\mu }D\left( {\phi ,\theta ,\gamma } \right) = \Lambda _{\text{\;}\nu}^{\mu}{\sigma ^\nu },
\end{equation}
with $D\left( {\phi ,\theta ,\gamma } \right)$  the spin-$\frac{1}{2}$ rotation matrix,
\begin{equation}
\Lambda _{\text{\;}0}^{0}=1,
\end{equation}
\begin{equation}
\Lambda _{\text{\;}0}^{i}=\Lambda _{\text{\;}j}^{0}=0
\end{equation}
and
\begin{equation}
\Lambda _{\text{\;}j}^{i}= {R_{ij}}\left( {\phi ,\theta ,\gamma } \right),
\end{equation}
with $R\left( {\phi ,\theta ,\gamma } \right)$ the spin-1 rotation matrix. If both the momentum of the polarization vector and the reference momentum are rotated, one can see that under rotations the chosen polarization vectors do transform as
\begin{equation}
\varepsilon _h^{*\mu }\left( {k,r} \right) \to \Lambda _{\text{\;}\nu}^{\mu} \varepsilon _h^{*\nu }\left( {k,r} \right)
\end{equation}
(due to the freedom in the reference momentum, it is possible to rotate both vectors simultaneously; by performing it in this way, the typical gauge term is not produced) \cite{dreiner}.

\smallskip
Also, direct use of (\ref{polr}) and (\ref{poll}) shows that the completeness relation is that of a light-like axial gauge,
\begin{equation}
\sum\limits_{h =  \pm } {\varepsilon _h^{*\mu }\left( {k,r} \right){{\left[ {\varepsilon _h^{*\nu }\left( {k,r} \right)} \right]}^*}}  =  - {g^{\mu \nu }} + \frac{{{k^\mu }{r^\nu } + {k^\nu }{r^\mu }}}{{k \cdot r}}.
\end{equation}
And finally, with the particular choices of vectors $k = \left( {{k_0},0,0,{k_0}} \right)$ and $r = \left( {{r_0},0,0, - {r_0}} \right)$, with	${k_0},{r_0} > 0$, up to a phase one may write
\begin{equation}
{u_L}\left( k \right) = \sqrt {2{k_0}} \left( {\begin{array}{*{20}{c}}
0\\
1
\end{array}} \right)
\end{equation}
and
\begin{equation}
{u_L}\left( r \right) = \sqrt {2{r_0}} \left( {\begin{array}{*{20}{c}}
{ - 1}\\
0
\end{array}} \right).
\end{equation}
When writing the respective polarization vector, one finds the familiar form for propagation in the $z$ direction,
\begin{equation}
\varepsilon _ + ^*\left( {k,r} \right) =  - \frac{1}{{\sqrt 2 }}{\left( {0,1,i,0} \right)^*}.
\end{equation}

\end{document}